\documentclass[english,10pt,letterpaper]{article}
\usepackage[T1]{fontenc}
\usepackage{graphicx}
\usepackage{mathtools}
\usepackage{amssymb}
\usepackage{amsthm}
\usepackage{xcolor}
\usepackage{babel}
\usepackage[hidelinks]{hyperref}
\usepackage{dsfont}
\usepackage{subfig}
\usepackage[left=2.4cm,right=2.4cm,top=2cm,bottom=2cm]{geometry}
\usepackage{orcidlink}
\usepackage{bm}
\usepackage{mwe}
\usepackage{physics}
\usepackage{slashed}
\usepackage{lmodern}
\usepackage{tikz}

\newcommand{\bk}[2]{\left\langle #1 | #2 \right\rangle}
\newcommand{\bkf}[3]{\left\langle #1 \right| #2 \left| #3 \right\rangle}
\newcommand{\pj}[2]{ \left| #1 \right\rangle \left\langle #2 \right|}

\newcommand{\Kp}{\underline{\otimes}}

\newcommand{\mfs}[1]{\left[ \! \left[ #1 \right] \! \right]}

\begin{document}
	
\title{{\bf $\mathcal{N}$-bein formalism for degenerate states in the parameter space of quantum geometry}}
	
\author{Jorge Romero \orcidlink{0000-0001-8258-6647}, Carlos A. Velasquez \orcidlink{0009-0003-1513-5098}, and J David Vergara \orcidlink{0000-0002-8615-761X} \footnote{Corresponding author} }
\date{}

\maketitle	

\vspace{-1cm}
\begin{center}
	\small
	Departamento de F\'{i}sica de Altas Energ\'{i}as, Instituto de Ciencias Nucleares, Universidad Nacional Aut\'{o}noma de M\'{e}xico, Apartado Postal 70-543, Ciudad de M\'{e}xico, 04510, Mexico.\\
\end{center}

\begin{flushleft}
	Email: jorge.romero@correo.nucleares.unam.mx, carlos.velasquez@correo.nucleares.unam.mx, and vergara@nucleares.unam.mx\\[0.5cm]
\end{flushleft}

\begin{abstract}
\noindent
Recently, we introduced a geometric object analogous to an orthonormal frame in the Cartan formalism to study the parameter space of quantum systems; we called it $\mathcal{N}$-bein, with $\mathcal{N}$ being the number of parameters that characterize the quantum system. Acting as the ``square root'' of the quantum geometric tensor (QGT), the $\mathcal{N}$-bein allows us to define new tensors to improve our understanding of the structure beneath the parameter space of quantum mechanics. In this work, we extend this mathematical framework surrounding the $\mathcal{N}$-bein to analyze the parameter space of quantum systems with degenerate spectra. As in the non-degenerate case, we define a non-Abelian two-state QGT to identify possible transitions between degenerate states after two consecutive parameter variations. Additionally, using the Wilczek-Zee connection, we introduce a torsion-like tensor as the covariant derivative of the $\mathcal{N}$-bein. This torsion captures the noncommutativity of successive parameter variations and coincides with the antisymmetric part of the two-state QGT. We also present a geometrical formulation using differential forms and discuss the physical implications of the newly defined tensors. Furthermore, we construct several gauge-invariant observables from the $\mathcal{N}$-bein and its derivatives to highlight the utility of the new tensors. Finally, to illustrate the convenience and applications of this formalism, we apply the theoretical framework to a system of coupled harmonic oscillators immersed in an electric field. The coupling between the oscillators results in a degenerate system. Thus, using the new formalism, we found correlations among the quantum states quantified by the new invariants.
\end{abstract}


\section{Introduction}
Geometric concepts have profoundly enriched physics, offering visual and conceptual insight into diverse phenomena. Their impact is evident in fields like General Relativity (GR) \cite{Straumann2012general} and the Standard Model (SM) \cite{exp2}, which explores fundamental forces. The development of geometric tools for studying quantum phenomena has increased substantially in recent years \cite{Torma}. These tools have proven relevant in quantum mechanics \cite{Braunstein}, condensed matter physics \cite{Bernevig}, and quantum optics \cite{Ezawa,Ghosh}. 

This geometric framework has been increasingly applied in quantum mechanics \cite{Provost, Berry}, particularly in the analysis of the parameter space of quantum systems \cite{Carollo}. In this field, the evolution of quantum states and their properties are examined within parameter spaces. The fundamental element in all these studies has been the quantum geometric tensor (QGT) \cite{Provost}, which contains the quantum metric as its symmetric part and the Berry curvature \cite{Berry} as its anti-symmetric part. The QGT quantifies how parameter variations affect quantum information encoded in states, and it is gauge invariant. 

 This tensor provides  a unified framework for describing the topology and geometry of quantum states in different reference spaces. In the parameter space, it defines the distance between quantum states \cite{Zanardi, Kolodrubetz, Hetenyi, Alvarez}; in momentum space, it identifies the topological phases of materials \cite{Bleu, Matsuura, Palumbo, Ahn, Northe, Graf, Bouhon, Kaplan}; and, in configuration space, it introduces a fidelity marker that distinguishes between different phases—such as metallic and insulating \cite{Marrazzo, deSousa}—or serves as an entanglement marker \cite{Juarez}. Moreover, recent studies demonstrate that the quantum metric induces non-linear conductivities that dominate the transport properties of PT-symmetric systems \cite{Agarwal1}.

Traditionally, these geometric quantities are described within Riemannian geometry. However, Cartan geometry provides an alternative and robust framework. Cartan's approach generalizes Riemannian geometry by introducing the vielbein and affine connection, rather than relying solely on the metric. These elements construct the Riemann curvature (connection curvature) and the torsion 2-form (covariant derivative of the vielbein). 
Although the standard formalism of the QGT breaks down in the presence of degenerate flat bands or identical energy levels, such systems can be effectively analyzed using non-Abelian frameworks like the Wilczek-Zee (WZ) curvature \cite{Wilczek}. Indeed, for degenerate systems, one can construct both the WZ connections and a generalized, non-Abelian QGT \cite{Yu-Quan, Rezakhani}. This extended formalism is highly versatile: it facilitates the description of symmetry breaking and quantum phase transitions in degenerate regimes. Furthermore, the associated quantum metric governs the error-minimizing trajectories in holonomic quantum computation, laying the groundwork for topological quantum computation \cite{Lahtinen}. In a broader topological context, the non-Abelian quantum metric has also been utilized to derive the real Chern number of a generalized Dirac monopole and the second Chern number of a Yang monopole \cite{Ding}.

In a previous publication, we introduced Cartan's geometry to describe the quantum parameter space~\cite{Romero2409}. This description introduced the $\mathcal{N}$-bein, which corresponds to the well-known non-adiabatic coupling vectors \cite{Tully, Crespo, Kryachko0300}. However, we reinterpreted them as the ``square root'' of the QGT—a shift that reveals the underlying geometric structure of the $\mathcal{N}$-bein and enables the construction of new tensors. Furthermore, a non-zero $\mathcal{N}$-bein indicates a possible change between two energy levels due to a variation in the parameter $\lambda^{i}$, where $i=1,\ldots,\mathcal{N}$. In contrast, the Berry connection only accounts for parameter variations that do not change the original state. Thus, our approach complements the existing framework. 

In this paper, we extend our analysis of Cartan geometry to degenerate systems,  as shown in section~\ref{Sec_e_def}. This scenario introduces several novel features that extend beyond standard non-Abelian gauge theories. Alongside the WZ connections, we must introduce a degenerate $\mathcal{N}$-bein, which takes the form of a rectangular matrix rather than the square matrices typical of WZ connections. This rectangular structure arises because the degenerate $\mathcal{N}$-beins must bridge two states that may have different degeneracies, resulting in distinct gauge transformations acting on the left and right sides of the $\mathcal{N}$-bein, as illustrated in equation \eqref{tranbein}. This important extension recognizes that adiabatic transport within degenerate subspaces yields matrix-valued transformations of quantum states, rather than merely scalar phase factors. Thus, degeneracy is not just a quantitative detail, but a fundamental structural feature that transforms the geometric phase framework from an Abelian (scalar) to a non-Abelian (matrix-valued) theory, unlocking a richer array of quantum phenomena such as topological order and fault-tolerant quantum computation. Furthermore, these $\mathcal{N}$-beins yield new gauge-invariant information: specifically, the trace of their squared moduli is gauge-invariant and provides detailed insights into possible transitions—both between states with the same degeneracy and those with different degeneracies. This will be explored in detail through the example in section 7. Another relevant result of this work is that, with the $\mathcal{N}$-beins, we derive a more efficient way to compute the non-Abelian QGT.

In section~\ref{Sec_tsQGT}, we extend the concept of the two-state geometric tensor—originally introduced in~\cite{Romero2409}—to the degenerate case. This tensor consists of a symmetric component \eqref{GyM} and an antisymmetric component \eqref{TyM}. Because these components are complex and gauge-dependent, they do not correspond to direct physical observables. Nevertheless, by utilizing the associated invariants defined in section~\ref{Sec_invs}, this tensor provides a framework for understanding the correlations between two quantum states with potentially different degeneracies, which become correlated via variations in two parameters.

Furthermore, in section~\ref{Sec_Cartan}, we define a covariant derivative associated with the degenerate space. It uses the WZ connection to maintain the covariance in the tensors. In particular, the covariant derivative of the $\mathcal{N}$-bein results in a torsion-like tensor, by analogy with Cartan's geometry \cite{KN1}. This torsion corresponds to the anti-symmetric part of the two-state QGT, similar to how the WZ curvature corresponds to the anti-symmetric part of the degenerate QGT. 

In section~\ref{Sec_invs}, we employ the tools developed in sections~\ref{Sec_e_def}, \ref{Sec_tsQGT}, and \ref{Sec_Cartan} to present new gauge-invariant quantities, i.e., the physical observables of our theory. Among the invariants, we found tensors with symmetries akin to those of the Riemann tensor, vectors that do not depend on the states' degeneracy, and scalars that provide a notion of norm for the tensors introduced in this formalism. 

In section~\ref{Sec_df}, we reformulate our formalism in terms of differential forms, providing a more precise geometrical interpretation. In particular, we obtain a 3-form invariant that has a reminiscence of a Chern-Simons 3-form \cite{Chern-Simons}. Thus, its introduction motivates a new topological invariant, analogous to the Nieh-Yan invariant \cite{Nieh1,Nieh}. This topological invariant has not been previously introduced in the field of condensed matter, and we expect it to be highly relevant for quantum systems that depend on three or more parameters.

In section~\ref{Sec_ejem}, we demonstrate the advantages of our formalism through a concrete example. In particular, we found all the possible state transitions induced by parameter variations. These results encompass transitions between distinct subspaces with different degeneracy, as well as transitions within a single subspace. Furthermore, we show that several of our invariants give us topological information in the sense that they only depend on the energy levels and are independent of the parameters. 

Finally, in section~\ref{Sec_concl}, we present the conclusions of our work. Two appendices complement the main article. Appendix~\ref{Ap_prop} compiles properties of the tensors and invariants used throughout the manuscript, while appendix~\ref{Ap_Ex_extra} offers further analysis of the example treated in section~\ref{Sec_ejem}.

\section{$\mathcal{N}$-bein definition}
\label{Sec_e_def}

Let $\mathcal{H}$ be the Hilbert space composed by the subspaces $\mathcal{H}_{n}$, $\mathcal{H} = \bigoplus\limits_{n} \mathcal{H}_{n}$, it is the orthogonal direct sum of the subspaces. Each subspace $\mathcal{H}_{n}$ is characterized by the energy $E_{n}$. Hence, the states $\ket{n_{N}}\in \mathcal{H}_{n}$ satisfy the Schr\"{o}dinger equation
\begin{equation}
	\label{Sch_eq}
	\hat{H} \ket{n_{N}} = E_{n} \ket{n_{N}}.
\end{equation}
We use $N=1,\dots, d_{n}$ to number the eigenstates with energy $E_{n}$, such that $d_{n}$ is the finite degeneracy in $\mathcal{H}_{n}$. Moreover, the label $N$ identifies a direction on the subspace $\mathcal{H}_{n}$. Thus, the set $\{\ket{n_{N}}\}_{N=1}^{d_{n}}$ forms a basis for $\mathcal{H}_{n}$, and $\{\{\ket{n_{N}}\}_{N=1}^{d_{n}}\}_{n=0}^{\infty}$ is the basis for the entire Hilbert space $\mathcal{H}$. Equivalently, each subspace $\mathcal{H}_{n}$ has its identity
\begin{equation}
\label{idn}
	\hat{\mathds{1}}^{(n)} = \sum_{N=1}^{d_{n}} \pj{n_{N}}{n_{N}}.
\end{equation}
Therefore, the resolution of the identity is
\begin{equation}
	\label{id}
	\hat{\mathds{1}} = \sum_{n=0}^{\infty} \hat{\mathds{1}}^{(n)}= \sum_{n=0}^{\infty}\sum_{N=1}^{d_{n}} \pj{n_{N}}{n_{N}}.
\end{equation}
Furthermore, for two arbitrary states, $\ket{n_{N}}$ and $\ket{m_{M}}$, the orthonormality condition reads
\begin{equation}
	\label{norm}
	\bk{m_{M}}{n_{N}} = \delta_{mn} \delta_{MN}.
\end{equation}
We use uppercase letters to indicate the degeneracy within the energy level specified by its lowercase counterpart. Thus, the indices $N, N_{1}, N_{2}, \dots$ label directions on $\mathcal{H}_{n}$, i.e., they represent different states with energy $E_{n}$. Similarly, for a different subspace $\mathcal{H}_{m}$, we use $M, M_{1}, M_{2}, \dots$ to label the states with energy $E_{m}$. This notation is convenient to distinguish the degeneracy of different energy levels. Because in general $d_{n} \neq d_{m}$, so the indices $N, N_{1}, N_{2}, \dots$ may take values different from $M, M_{1}, M_{2}, \dots$.

Now that we have defined the Hilbert space, we consider that each state $\ket{n_{N}} \in \mathcal{H}_{n} \subset \mathcal{H} $ depends on the $\mathcal{N}$ parameters $\lambda=\{\lambda^{i}|i=1,\dots,\mathcal{N}\}$, where each $\lambda^{i}$ is a slowly varying function of time. Thus, $\ket{n_{N}}= \ket{n_{N}(\lambda)}$. Although all the states considered in this paper depend on $\lambda$, for simplicity, we often omit writing its dependency. A small variation on the parameters $\lambda^{i} \rightarrow \lambda^{i} + \delta \lambda^{i}$ modifies the state
\begin{equation}
	\ket{n_{N}(\lambda + \delta \lambda)} = \ket{n_{N}(\lambda)} + \ket{\partial_{i} n_{N} } \delta \lambda^{i} + \dots,
\end{equation}
with $\ket{\partial_{i} n_{N}} := \frac{\partial}{\partial \lambda^{i}} \ket{n_{N}(\lambda)}$. Therefore, a perturbation on the parameters changes the state. At first order, we have
\begin{eqnarray}
	\ket{\delta n_{N}} & := & \ket{n_{N}(\lambda + \delta \lambda)} - \ket{n_{N}(\lambda)} \approx \ket{\partial_{i} n_{N}} \delta \lambda^{i}\nonumber \\
	\label{expansion}
	& = & \left( \sum_{N_{1}=1}^{d_{n}} \bk{n_{N_{1}}}{\partial_{i} n_{N}} \ket{n_{N_{1}}} + \sum_{m \neq n}\sum_{M=1}^{d_{m}} \bk{m_{M}}{\partial_{i} n_{N}} \ket{m_{M}} \right) \delta \lambda^{i}, 
\end{eqnarray}
where in the last equality we used the resolution to the identity \eqref{id} to write $\ket{\partial_{i} n_{N}}$ in the unperturbed basis. Also, we are considering Einstein's sum convention for the indices labeling the parameters.

The first coefficient in the expansion is proportional to the WZ connection, 
\begin{equation}
	\label{A}
	A^{(n)}_{i \, N_{1} N_{2}} := \mathrm{i} \bk{n_{N_{1}}}{\partial_{i} n_{N_{2}}},
\end{equation}
which indicates the part of the change that stays in the subspace $\mathcal{H}_{n}$ after a variation on $\lambda^{i}$. The different indices $N_{1}$ and $N_{2}$ illustrate a possible direction shift within the subspace $\mathcal{H}_{n}$.  Meanwhile, the second part of~\eqref{expansion} accounts for a potential transition to a different subspace $\mathcal{H}_{m}$, where the indices $N$ and $M$ are the directions in the former and latter subspaces, respectively. Following the guidance of~\cite{Romero2409}, we define the degenerate $\mathcal{N}$-bein as
\begin{equation}
	\label{e_def}
	e^{(m,n)}_{i\; MN} := \mathrm{i} \bk{m_{M}}{\partial_{i} n_{N}}. 
\end{equation}	
The $\mathcal{N}$-bein, as in~\cite{Romero2409}, will be our key component; with it, we construct all the important quantities to study the parameter space of degenerate systems. 

Similar to the work of Zanardi~\cite{Zanardi, Zanardi0708}, using equation~\eqref{Sch_eq}, we derive the equivalent formula for the $\mathcal{N}$-bein:
\begin{equation}
	\label{e_def2}
	e^{(m,n)}_{i\; MN} = \mathrm{i} \dfrac{\bkf{m_{M}}{\partial_{i} \hat{H}}{n_{N}}}{E_{n} - E_{m}}.
\end{equation}
Both expressions for $e^{(m,n)}_{i\; MN}$ are consistent with the work in~\cite{Romero2409} when there is no degeneracy, $d_{n}=1$ for all $n$. In fact, all subsequent geometric objects derived here will reduce to those in~\cite{Romero2409} when we take the limit of no degeneracy. Additionally, notice that the indices $M=1,\dots, d_{m}$ and $N=1,\dots, d_{n}$ label the components of the $\mathcal{N}$-bein as the elements of a $d_{m} \times d_{n}$ rectangular matrix, $M$ indicates the row and $N$ the column. Thus,  each entry looks like a non-degenerate $\mathcal{N}$-bein $e^{(m,n)}_{i}$. This feature is an extension of the usual formalism of non-Abelian theories, where, as in the case of WZ connections, they are always associated with square matrices. However, as we will see in section~\ref{Sec_Cartan}, the transformation rules of the non-Abelian $\mathcal{N}$-bein are perfectly well defined; see \eqref{tranbein}. 

\section{Two-state QGT}
\label{Sec_tsQGT}	

For degenerate systems, the QGT is~\cite{Yu-Quan}
\begin{equation}
	Q^{(n)}_{ij\; N_{1} N_{2}} := \bra{\partial_{i}n_{N_{1}}} \left( \hat{\mathds{1}} - \sum_{N=1}^{d_{n}} \pj{n_{N}}{n_{N}} \right) \ket{\partial_{j} n_{N_{2}}}.
\end{equation}
Thus, using~\eqref{id} and \eqref{e_def}, it is straightforward to arrive at the expression
\begin{equation}\label{qgt_def}
	Q^{(n)}_{ij\; N_{1} N_{2}} = \sum_{m \neq n} \sum_{M=1}^{d_{m}} e^{(m, n)\, \ast}_{i\; M N_{1}} e^{(m,n)}_{j\; M N_{2}},
\end{equation}
where the asterisk denotes complex conjugation. Therefore, $e^{(m,n)}_{i\; MN}$ behaves as the ``square root'' of the QGT. Furthermore, using the property \eqref{conj}, we write the QGT as 
\begin{align}
	\label{Q_alt}
	Q^{(n)}_{ij\; N_{1} N_{2}}& = \sum_{m \neq n} \sum_{M=1}^{d_{m}} e^{(n,m)}_{i\; N_{1} M} e^{(m,n)}_{j\; MN_{2}}  = - \sum_{m \neq n} \sum_{M=1}^{d_{m}} \bk{n_{N_{1}}}{\partial_{i} m_{M}} \bk{m_{M}}{\partial_{j} n_{N_{2}}} .
\end{align}
In this form, we reinterpret $Q^{(n)}_{ij\; N_{1} N_{2}}$ as the quantity that encodes the transition from one degenerate state to another within the same subspace $\mathcal{H}_{n}$ after two consecutive parameter variations.  The sums in ~\eqref{Q_alt} account for all the possible paths that leave $\mathcal{H}_{n}$ from $\ket{n_{N_{1}}}$ to $\ket{n_{N_{2}}}$ or vice versa. When $Q^{(n)}_{ij\; N_{1} N_{2}}$ is symmetric in the parameter indices, the order of the parameter variations does not matter; otherwise, it does. Hence, we separate this tensor into its symmetric and anti-symmetric parts. The splitting results in two matrix-valued tensors: the non-Abelian metric for the parameter space and the curvature of the WZ connection; respectively, they are:
\begin{subequations}
	\begin{eqnarray}
        g^{(n)}_{ij\; N_{1} N_{2}} &:=& \mbox{Sym}\left( Q^{(n)}_{ij\; N_{1} N_{2}} \right)  = \frac{1}{2} Q^{(n)}_{ij\; N_{1} N_{2}} + \frac{1}{2}     Q^{(n)}_{ji\; N_{1} N_{2}} \nonumber\\
		      &=&  \frac{1}{2} \sum_{m \neq n} \sum_{M=1}^{d_{m}} \Big( e^{(n,m)}_{i\; N_{1} M} e^{(m,n)}_{j\; MN_{2}}
        + e^{(n,m)}_{j\; N_{1} M} e^{(m,n)}_{i\; MN_{2}}\Big),  \\
		F^{(n)}_{ij\; N_{1} N_{2}} &:=& 2 \mathrm{i} \mbox{ASym}\left( Q^{(n)}_{ij\; N_{1} N_{2}} \right) =  \mathrm{i} Q^{(n)}_{ij\; N_{1} N_{2}} - \mathrm{i} Q^{(n)}_{ji\; N_{1} N_{2}}  \nonumber \\
		      &=& \mathrm{i} \sum_{m \neq n} \sum_{M=1}^{d_{m}} \Big( e^{(n,m)}_{i\; N_{1} M} e^{(m,n)}_{j\; MN_{2}}
         - e^{(n,m)}_{j\; N_{1} M} e^{(m,n)}_{i\; MN_{2}} \Big).
	\end{eqnarray}
\end{subequations}

From the latter interpretation of the QGT, we ponder the possibility of studying the change from one subspace $\mathcal{H}_{n}$ to a different one $\mathcal{H}_{m}$ after consecutive parameter variations. Thus, we define the two-state QGT for degenerate systems as
\begin{align}
	\label{M_def}
	M^{(m,n)}_{ij\; MN}& := \sum_{l \neq n,m} \sum_{L=1}^{d_{l}} e^{(l,m)\, \ast}_{i\; LM} e^{(l,n)}_{j\; LN} = \sum_{l \neq n,m} \sum_{L=1}^{d_{l}}  e^{(m,l)}_{i\; ML} e^{(l,n)}_{j\; LN}.
\end{align}
Again, we are summing over all the possible states that connect $\ket{m_{M}}$ and $\ket{n_{N}}$  after the variations of $\lambda^{i}$ and $\lambda^{j}$.

Looking closely at the last equality in~\eqref{M_def}, we see an $\mathcal{N}$-bein that encodes the transition from the state $\ket{n_{N}}$ to $\ket{l_{L}}$ when the $j$th parameter is varied. In the same way, the other $\mathcal{N}$-bein possesses information on the transition from $\ket{l_{L}}$ to $\ket{m_{M}}$ after a variation of the $i$th parameter. Taken together, this means that the two-state QGT identifies a transition from $\ket{n_{N}}$ to $\ket{m_{M}}$, with the sum capturing all the available intermediate states $\ket{l_{L}}$ that mediate this connection and that are outside $\mathcal{H}_{n}$ and $\mathcal{H}_{m}$.

Moreover, the symmetries on $M^{(m,n)}_{ij\; MN}$ provide fundamental information about the transitions. When the tensor is symmetric, the order of parameter variations does not matter, but when it is anti-symmetric, the order makes a difference. Therefore, we divide the new tensor into its symmetric and anti-symmetric parts:
\begin{subequations}
	\begin{eqnarray}
		\mathcal{G}^{(m,n)}_{ij\; MN} &:=& \mbox{Sym}\left( M^{(m,n)}_{ij\; MN} \right) = \frac{1}{2} M^{(m,n)}_{ij\; MN} + \frac{1}{2} M^{(m,n)}_{ji\; MN} \nonumber \\ &=& \frac{1}{2} \sum_{l \neq n,m} \sum_{L=1}^{d_{l}} \left( e^{(m,l)}_{i\; ML} e^{(l,n)}_{j\; LN} + e^{(m,l)}_{j\; ML} e^{(l,n)}_{i\; LN} \right),  \label{GyM} \\
		T^{(m,n)}_{ij\; MN} &:=& 2 \mathrm{i} \mbox{ASym}\left( M^{(m,n)}_{ij\; MN} \right) = \mathrm{i} M^{(m,n)}_{ij\; MN} - \mathrm{i} M^{(m,n)}_{ji\; MN}\nonumber \\ &=& \mathrm{i} \sum_{l \neq n,m} \sum_{L=1}^{d_{l}} \left( e^{(m,l)}_{i\; ML} e^{(l,n)}_{j\; LN} - e^{(m,l)}_{j\; ML} e^{(l,n)}_{i\; LN} \right)  \label{TyM}.
	\end{eqnarray}
\end{subequations}
The tensors $\mathcal{G}^{(m,n)}_{ij\; MN}$ and $T^{(m,n)}_{ij\; MN}$ tell us, in precise quantitative terms, whether the order of variations we take through the parameter space changes the outcome of the transition from $\ket{n_{N}}$ to $\ket{m_{M}}$.	

It is important to emphasize that, by construction, we are only considering transitions from one subspace $\mathcal{H}_{n}$ to another $\mathcal{H}_{m}$, with $n \neq m$. Hence, $M^{(m,n)}_{ij\; MN}$ is not a generalization of $Q^{(n)}_{ij\; N_{1} N_{2}}$, rather it is an additional tool to study the parameter space of degenerate systems. Furthermore, keep in mind that the tensors $\mathcal{G}^{(m,n)}_{ij\; MN}$ and $T^{(m,n)}_{ij\; MN}$ are not necessarily real, they are complex matrix-valued tensors. Nevertheless, we use them to construct real gauge-invariant quantities as observables. We elaborate on this idea in section~\ref{Sec_invs}.

\section{Gauge transformations and torsion}
\label{Sec_Cartan}	

Consider a state $|n_N\rangle$ belonging to the Hilbert subspace $\mathcal{H}_n$ with degeneracy $d_n$. This state transforms as 
\begin{equation}
	\label{gauge}
	\ket{n_{N}}^{\prime} = \sum_{N_{1}=1}^{d_{n}}  \ket{n_{N_{1}}} U^{(n)}_{N_{1} N},
\end{equation}	
under a gauge transformation, where $U^{(n)}_{N_{1} N_{2}}$ are the components of a unitary $d_{n} \times d_{n}$ matrix $\mathds{U}^{(n)}$, $\mathds{U}^{(n)\, \dagger} \mathds{U}^{(n)}= \mathds{U}^{(n)} \mathds{U}^{(n)\, \dagger} = \mathds{1}^{(n)}$. This transformation maintains the norm of a state $\ket{n_{N}}$ because $\mathds{U}^{(n)}$ is unitary,
\begin{equation}
	\left(\bk{n_{N_{1}}}{n_{N_{2}}}\right)^{\prime} = \bk{n_{N_{1}}}{n_{N_{2}}} = \delta_{N_{1} N_{2}}.
\end{equation}
Equation~\eqref{gauge} defines a change of basis within $\mathcal{H}_{n}$. Generally, the transformation matrix is a function of the parameters, $U^{(n)}_{N_{1} N_{2}} = U^{(n)}_{N_{1} N_{2}} (\lambda)$. This transformation was studied by Wilczek and Zee \cite{Wilczek}, who noted that the object $A^{(n)}_{i\; N_{1} N_{2}}$, defined in~\eqref{A}, transforms as a non-Abelian connection:
\begin{align}
	&\left( A^{(n)}_{i \, N_{1} N_{2}} \right)^{\prime} =
	 \sum_{N_{3}=1}^{d_{n}} \Big( \mathrm{i} U^{(n)\, \ast}_{N_{3} N_{1}} \partial_{i} U^{(n)}_{N_{3} N_{2}} +  \sum_{N_{4}=1}^{d_{n}} U^{(n) \,\ast}_{N_{3} N_{1}} A^{(n)}_{i\, N_{3} N_{4}} U^{(n)}_{N_{4} N_{2}} \Big).
\end{align}
Hence, its curvature is
\begin{equation}
	\label{F}
	F^{(n)}_{ij \; N_{1} N_{2}} := \partial_{i} A^{(n)}_{j\; N_{1} N_{2}} - \partial_{j} A^{(n)}_{i\; N_{1} N_{2}} - \mathrm{i} \sum_{N_{3}=1}^{d_{n}} \left( A^{(n)}_{i \; N_{1} N_{3}} A^{(n)}_{j \; N_{3} N_{2}} -  A^{(n)}_{j \; N_{1} N_{3}} A^{(n)}_{i \; N_{3} N_{2}} \right),
\end{equation}
which is also the anti-symmetric part of $Q^{(n)}_{ij\; N_{1} N_{2}}$. This curvature transforms as 
\begin{equation}
	\left( F^{(n)}_{ij \; N_{1} N_{2}} \right)^{\prime} = \sum_{N_{3}=1}^{d_{n}} \sum_{N_{4}=1}^{d_{n}} U^{(n)\, \ast}_{N_{3} N_{1}} F^{(n)}_{ij \; N_{3} N_{4}}  U^{(n)}_{N_{4} N_{2}}.
\end{equation}

Meanwhile, under the same transformation, the $\mathcal{N}$-bein transforms as
\begin{equation}
	\left( e^{(m,n)}_{i\; MN} \right)^{\prime} = \sum_{M_{1}=1}^{d_{m}} \sum_{N_{1}=1}^{d_{n}} U^{(m)\, \ast}_{M_{1} M} e^{(m,n)}_{i\; M_{1} N_{1}}  U^{(n)}_{N_{1} N}.\label{tranbein}
\end{equation}
In contrast with the transformation law for the WZ connection, the transformation for the $\mathcal{N}$-bein lacks the term with the partial derivative. Consequently, under this transformation, the $\mathcal{N}$-bein does not qualify as a connection, and we cannot define a curvature from it. Nonetheless, we use it to construct another important tensor. In Cartan's geometry, the torsion of a connection is related to the covariant derivative of the vielbein. Thus, we construct a torsion-like tensor using the $\mathcal{N}$-bein. Since the partial derivative alone does not preserve the transformation law, it cannot be used as a covariant derivative. Nonetheless, we found that the derivative
\begin{equation}
	D_{i} e^{(m,n)}_{j\; MN} := \partial_{i} e^{(m,n)}_{j\; MN} + \mathrm{i} \sum_{N_{1}=1}^{d_{n}} A^{(n)}_{i\; N_{1}N} e^{(m,n)}_{j\; M N_{1}} - \mathrm{i} \sum_{M_{1}=1}^{d_{m}} A^{(m)}_{i\; M M_{1}} e^{(m,n)}_{j\; M_{1}N}
\end{equation}
transforms as the $\mathcal{N}$-bein under the change of basis given by \eqref{gauge},
\begin{equation}
	\left( D_{i} e^{(m,n)}_{j\; MN} \right)^{\prime} = \sum_{M_{1}=1}^{d_{m}} \sum_{N_{1}=1}^{d_{n}} U^{(m)\, \ast}_{M_{1} M} D_{i} e^{(m,n)}_{j\; M_{1} N_{1}}  U^{(n)}_{N_{1} N}.
\end{equation}
Therefore, the WZ connection allows us to compute a covariant derivative for multiple-index objects. Notice how the connection is associated with the corresponding subspace of the index. A connection $A^{(n)}_{i\; N_{1} N_{2}}$, formed with the states of $\mathcal{H}_{n}$, transforms the indices within ($N,\, N_{1}, \, N_{2}, \dots$), while $A^{(m)}_{i\; M_{1} M_{2}}$ transforms the $M$-indices. 

Given the covariant derivative, the torsion\footnote{We drop the ``-like'' suffix for simplicity.} is the anti-symmetric part, because the torsion is an anti-symmetric $\binom{0}{2}$-type tensor, or 2-form. Explicitly,
\begin{eqnarray}
	\label{T_De}
	T^{(m,n)}_{ij\; MN} &:=& D_{i} e^{(m,n)}_{j\; MN} - D_{j} e^{(m,n)}_{i\; MN} \nonumber \\
	&=& \partial_{i} e^{(m,n)}_{j\; MN} - \partial_{j} e^{(m,n)}_{i\; MN} + \mathrm{i} \sum_{N_{1}=1}^{d_{n}} \left(  A^{(n)}_{i\; N_{1} N} e^{(m,n)}_{j\; M N_{1}} - A^{(n)}_{j\; N_{1} N} e^{(m,n)}_{i\; M N_{1}} \right) \nonumber \\
	&& - \mathrm{i} \sum_{M_{1}=1}^{d_{m}} \left( A^{(m)}_{i\; M M_{1}} e^{(m,n)}_{j\; M_{1}N} -A^{(m)}_{j\; M M_{1}} e^{(m,n)}_{i\; M_{1}N} \right).
\end{eqnarray}
Just as the curvature of the WZ connection is the anti-symmetric part of the QGT $Q^{(n)}_{ij\; N_{1} N_{2}}$, the torsion is the anti-symmetric part of the two-state QGT $M^{(m,n)}_{ij\; MN}$. To prove it, first we substitute the definitions \eqref{A} and \eqref{e_def} into \eqref{T_De}: 
\begin{eqnarray}
    T^{(m,n)}_{ij\; MN} &=&\text{i}\,\partial_{i}\left( \bk{m_{M}}{\partial_{j}n_{N}}\right) -\text{i}\,\partial_{j}\left( \bk{m_{M}}{\partial_{i}n_{N}}\right)  - \mathrm{i} \sum_{N_{1}=1}^{d_{n}}\bigg[   \bk{n_{N_{1}}}{\partial_{i}n_{N}}  \bk{m_{M}}{\partial_{j}n_{N_{1}}} \nonumber\\
    &&- \bk{n_{N_{1}}}{\partial_{j}n_{N}} \bk{m_{M}}{\partial_{i}n_{N_{1}}}\bigg] + \text{i} \sum_{M_{1}=1}^{d_{m}}\bigg[  \bk{m_{M}}{\partial_{i}m_{M_{1}}} \bk{m_{M_{1}}}{\partial_{j}n_{N}} \nonumber \\
    &&-\bk{m_{M}}{\partial_{j}m_{M_{1}}} \bk{m_{M_{1}}}{\partial_{i}n_{N}}\bigg].
\end{eqnarray}
Then, we compute the derivatives and rearrange the terms using~\eqref{e_norm_id} to identify the identity operators, see \eqref{idn} and \eqref{id}:
\begin{eqnarray}
    T^{(m,n)}_{ij\; MN} &=& \text{i}\bk{\partial_{i}m_{M}}{\partial_{j}n_{N}} -\text{i}  \bk{\partial_{j}m_{M}}{\partial_{i}n_{N}} + \mathrm{i} \bigg[\bra{\partial_{j}m_{M}}\left(\sum_{N_{1}=1}^{d_{n}}\ket{n_{N_{1}}}\bra{n_{N_{1}}}\right)\ket{\partial_{i}n_{N}} \nonumber\\
     &&- \bra{\partial_{i}m_{M}}\left(\sum_{N_{1}=1}^{d_{n}}\ket{n_{N_{1}}} \bra{n_{N_{1}}}\right)\ket{\partial_{j}n_{N}}\bigg] - \text{i}  \bigg[ \bra{\partial_{i}m_{M}}\left( \sum_{M_{1}=1}^{d_{m}}\ket{m_{M_{1}}} \bra{m_{M_{1}}}\right)\ket{\partial_{j}n_{N}} \nonumber \\
	&&  -\bra{\partial_{j}m_{M}}\left( \sum_{M_{1}=1}^{d_{m}}\ket{m_{M_{1}}} \bra{m_{M_{1}}}\right)\ket{\partial_{i}n_{N}} \bigg] \nonumber\\
    &=& \text{i} \bra{\partial_{i}m_{M}}\left(\hat{\mathds{1}}-\hat{\mathds{1}}^{(n)}-\hat{\mathds{1}}^{(m)}\right)\ket{\partial_{j}n_{N}} - \text{i} \bra{\partial_{j}m_{M}}\left(\hat{\mathds{1}}-\hat{\mathds{1}}^{(n)}-\hat{\mathds{1}}^{(m)}\right)\ket{\partial_{i}n_{N}}. 
\end{eqnarray}
From here, we use the identity
\begin{equation}
    \hat{\mathds{1}}-\hat{\mathds{1}}^{(n)}-\hat{\mathds{1}}^{(m)} \equiv \sum_{l \neq n,m} \sum_{L=1}^{d_{l}} \pj{l_L}{l_L},
\end{equation}
and we substitute it back into the torsion. After some simplification, we rewrite using \eqref{e_def} and \eqref{e_norm_id}:
\begin{eqnarray}
    T^{(m,n)}_{ij\; MN} &=& \text{i}\sum_{l\neq n,m}\sum_{L=1}^{d_{l}}\left( \bk{\partial_{i}m_{M}}{l_{L}} \bk{l_{L}}{\partial_{j}n_{N}} -  \bk{\partial_{j}m_{M}}{l_{L}}\bk{l_{L}}{\partial_{i}n_{N}}\right) \nonumber\\
    &=&\mathrm{i} \sum_{l \neq n,m} \sum_{L=1}^{d_{l}} \left( e^{(m,l)}_{i\; ML} e^{(l,n)}_{j\; LN} - e^{(m,l)}_{j\; ML} e^{(l,n)}_{i\; LN} \right).
\end{eqnarray}
Thus, we arrive at the equation \eqref{TyM}, proving the assertion.

It is important to emphasize that the gauge transformation only acts on the degeneracy indices, namely the uppercase letters ($M, N, \dots$). Therefore, taking these indices as the components of matrices and using matrix notation, the transformation laws are expressed in a more familiar form
\begin{equation}
	\left( A^{(n)}_{i} \right)^{\prime} =  \mathrm{i} \mathds{U}^{(n) \, \dagger} \partial_{i} \mathds{U}^{(n)} +  \mathds{U}^{(n) \,\dagger} A^{(n)}_{i} \mathds{U}^{(n)}.
\end{equation}
Similarly, the transformation law for its curvature is
\begin{equation}
	\left( F^{(n)}_{ij} \right)^{\prime} = \mathds{U}^{(n)\, \dagger}  F^{(n)}_{ij} \mathds{U}^{(n)}.
\end{equation}
On the other hand, for the rectangular matrices, the $\mathcal{N}$-bein and the torsion, the transformation laws are
\begin{subequations}
	\begin{eqnarray}
		\left( e^{(m,n)}_{i} \right)^{\prime} &=& \mathds{U}^{(m)\, \dagger} e^{(m,n)}_{i}  \mathds{U}^{(n)}, \\
		\left( T^{(m,n)}_{ij} \right)^{\prime} &=& \mathds{U}^{(m)\, \dagger} T^{(m,n)}_{ij}  \mathds{U}^{(n)}. 
	\end{eqnarray}
\end{subequations}

The transformation law for the curvature corresponds to a change of basis for a square matrix, so the matrices on its left and right are $d_{n}\times d_{n}$ unitary matrices. On the other hand, the $\mathcal{N}$-bein and the torsion have a rectangular matrix structure because the subspaces $\mathcal{H}_{m}$ and $\mathcal{H}_{n}$ could have different degeneracy. Therefore, their transformation laws require a $d_{m} \times d_{m}$ unitary matrix on the left and a $d_{n} \times d_{n}$ unitary matrix on the right. Nevertheless, the structure of their transformation is the same as that for the curvature.

\section{Gauge invariants}
\label{Sec_invs}

The tensors we have shown are not gauge invariant under the transformation defined in~\eqref{gauge}.  To extract measurable information from these geometric objects, we must use them to construct observables, which are real gauge invariants. For instance, the norm of a state $\ket{n_{N}}$ remains unchanged under a basis transformation. Although the tensors we defined transform as rectangular matrices under a basis transformation (they are not invariant individually), we build square matrices whose trace remains invariant under such transformations.

The first invariant we consider is the trace of the QGT
\begin{equation}
	\label{Qij_def}
	Q^{(n)}_{ij} := \sum_{N=1}^{d_{n}} Q^{(n)}_{ij\; NN} = \sum_{N=1}^{d_{n}} \sum_{m \neq n} \sum_{M=1}^{d_{m}} e^{(n,m)}_{i\; NM} e^{(m,n)}_{j\; MN},
\end{equation}	
as well as its symmetric and anti-symmetric parts:
\begin{subequations}
	\begin{eqnarray}
		\label{gij_def}
		g^{(n)}_{ij} &:=& \sum_{N=1}^{d_{n}} g^{(n)}_{ij\; NN} = \frac{1}{2} \sum_{N=1}^{d_{n}} \sum_{m \neq n} \sum_{M=1}^{d_{m}} \left( e^{(n,m)}_{i\; NM} e^{(m,n)}_{j\; MN} + e^{(n,m)}_{j\; NM} e^{(m,n)}_{i\; MN} \right),  \\
		\label{Fij_def}
		F^{(n)}_{ij} &:=& \sum_{N=1}^{d_{n}} F^{(n)}_{ij\; NN} = \mathrm{i} \sum_{N=1}^{d_{n}} \sum_{m \neq n} \sum_{M=1}^{d_{m}} \left(  e^{(n,m)}_{i\; NM} e^{(m,n)}_{j\; MN} - e^{(n,m)}_{j\; NM} e^{(m,n)}_{i\; MN} \right).
	\end{eqnarray}
\end{subequations}
Furthermore, the summands of \eqref{Qij_def} are also gauge invariants with useful applications (see~\cite{Ahn2203} for a non-degenerate example), explicitly:
\begin{equation}
	\label{qij_def}
	q^{(n,m)}_{ij} := \sum_{N=1}^{d_{n}} \sum_{M=1}^{d_{m}} e^{(n,m)}_{i\; NM} e^{(m,n)}_{j\; MN}.
\end{equation}	
Hence, when we sum over all energy levels $m\neq n$, we obtain the trace of the QGT
\begin{equation}
	\label{Qij_qij}
	Q^{(n)}_{ij} =  \sum_{m \neq n} q^{(n,m)}_{ij}.
\end{equation}
Although all of these quantities are invariant, only \eqref{gij_def} and \eqref{Fij_def} are considered observables because they are real tensors. 

The use of $\mathcal{N}$-beins is not required to formulate the previous invariants, but its use simplifies the computations. Meanwhile, for the next set of invariants, we use the new tensors defined throughout this work. First, we have the rank-3 and rank-4 tensor\footnote{We could also have defined an invariant with the $\mathcal{N}$-bein multiplying $\Xi^{(m,n)}_{ij\; MN}$ by the left. However, this invariant is proportional to $\left( R_{\Xi} \right)^{(m,n)}_{ijk}$.}:
\begin{eqnarray}
	\label{inv_R}
	\left( R_{\Xi} \right)^{(m,n)}_{ijk} &:=&  \sum_{N=1}^{d_{n}} \sum_{M=1}^{d_{m}} \Xi^{(m,n)}_{ij\; MN} e^{(n,m)}_{k\; NM}, \\
	\label{inv_S}
	\left( S_{\Xi \Theta} \right)^{(m,n)}_{ijkl} &:=& \sum_{N=1}^{d_{n}} \sum_{M=1}^{d_{m}} \Xi^{(m,n)}_{ij\; MN} \Theta^{(n,m)}_{kl\; NM}, 
\end{eqnarray}
where $\Xi^{(m,n)}_{ij\; MN}$ and $\Theta^{(m,n)}_{ij\; MN}$ could be any of the two-state tensors defined ($M^{(m,n)}_{ij\; MN}$, $\mathcal{G}^{(m,n)}_{ij\; MN}$, or $T^{(m,n)}_{ij\; MN}$).
Both $\left( R_{\Xi} \right)^{(m,n)}_{ijk}$ and $\left( S_{\Xi \Theta} \right)^{(m,n)}_{ijkl}$ are in general complex tensors, so the observables are derived from their real and imaginary parts. In particular, the real and imaginary parts of $\left( S_{\Xi \Theta} \right)^{(m,n)}_{ijkl}$ are
\begin{subequations}
	\begin{eqnarray}
		\left( N_{\Xi \Theta} \right)^{(m,n)}_{ijkl} &:=& \mathrm{Re} \left\lbrace \left( S_{\Xi \Theta} \right)^{(m,n)}_{ijkl}  \right\rbrace \nonumber \\
		&=&  \sum_{N=1}^{d_{n}} \sum_{M=1}^{d_{m}} \left( \mathrm{Re} \left\lbrace \Xi^{(m,n)}_{ij\; MN} \right\rbrace \mathrm{Re} \left\lbrace \Theta^{(n,m)}_{kl\; NM} \right\rbrace - \mathrm{Im} \left\lbrace \Xi^{(m,n)}_{ij\; MN} \right\rbrace \mathrm{Im} \left\lbrace \Theta^{(n,m)}_{kl\; NM} \right\rbrace \right) , \\
		\left( A_{\Xi \Theta} \right)^{(m,n)}_{ijkl} &:=& \mathrm{Im} \left\lbrace \left( S_{\Xi \Theta} \right)^{(m,n)}_{ijkl}  \right\rbrace \nonumber \\
		&=&  \sum_{N=1}^{d_{n}} \sum_{M=1}^{d_{m}} \left( \mathrm{Re} \left\lbrace \Xi^{(m,n)}_{ij\; MN} \right\rbrace \mathrm{Im} \left\lbrace \Theta^{(n,m)}_{kl\; NM} \right\rbrace + \mathrm{Im} \left\lbrace \Xi^{(m,n)}_{ij\; MN} \right\rbrace \mathrm{Re} \left\lbrace \Theta^{(n,m)}_{kl\; NM} \right\rbrace \right) .
	\end{eqnarray}
\end{subequations}
In the limit of no degeneracy ($d_{n}=1$, for all $n$), these observables are equivalent to those introduced in~\cite{Romero2409}, see discussion at the end of appendix~\ref{Ap_prop}.

Since the metric $g^{(n)}_{ij}$ is a real invariant, its determinant will be an observable as well. Moreover, when $\det(g^{(n)}_{ij})\neq 0$, we use the inverse metric $g^{(n)\, ij}$ to define a vector and a scalar invariants
\begin{subequations}
	\begin{eqnarray}
		\label{Inv_vec}
		\Xi^{(m,n)\, i} &:=& g^{(m)\, ij} g^{(n)\, kl} \left( R_{\Xi} \right)^{(m,n)}_{jkl}  = \sum_{N=1}^{d_{n}} \sum_{M=1}^{d_{m}} g^{(m)\, ij} \Xi^{(m,n)}_{jk\; MN} g^{(n)\, kl} e^{(n,m)}_{l\; NM}, \\
		\label{Inv_scl}
		\mathcal{N}_{\Xi \Theta}^{(m,n)} &:=& 2 g^{(m)\, ij} g^{(n)\, kl} \left( S_{\Xi \Theta} \right)^{(m,n)}_{jkli} =2 \sum_{N=1}^{d_{n}} \sum_{M=1}^{d_{m}} g^{(m)\, ij} \Xi^{(m,n)}_{jk\; MN}  g^{(n)\, kl} \Theta^{(n,m)}_{li\; NM}.
	\end{eqnarray}
\end{subequations}
For the scalar invariant, we have the case 
\begin{equation}
	\label{Inv_NX}
	\mathcal{N}_{\Xi}^{(m,n)} := \mathcal{N}_{\Xi \Xi}^{(m,n)} = 2 \sum_{N=1}^{d_{n}} \sum_{M=1}^{d_{m}} g^{(m)\, ij} \Xi^{(m,n)}_{jk\; MN}  g^{(n)\, kl} \Xi^{(n,m)}_{li\; NM},
\end{equation}
which is always a real quantity. Thus, this observable is like a norm for any of the tensors defined ($M^{(m,n)}_{ij\; MN}$, $\mathcal{G}^{(m,n)}_{ij\; MN}$, or $T^{(m,n)}_{ij\; MN}$), and it is symmetric to the interchange of states
\begin{equation}
	\label{inv_sym}
	\mathcal{N}_{\Xi}^{(m,n)}=\mathcal{N}_{\Xi}^{(n,m)}.
\end{equation}
Therefore, for a given tensor $\Xi^{(m,n)}_{ij\; MN}$, $\mathcal{N}_{\Xi}^{(m,n)}$ measures its magnitude between the subspaces  $\mathcal{H}_{m}$ and $\mathcal{H}_{n}$. This invariant reduces to the one in~\cite{Romero2409}. On the other hand, invariants $\left( R_{\Xi} \right)^{(m,n)}_{ijk}$ and $\Xi^{(n,m)\; i}$ do not have a counterpart in such a reference. However, we easily derive their corresponding limit for the non-degenerate case as
\begin{subequations}
	\begin{eqnarray}
		\left( R_{\Xi} \right)^{(m,n)}_{ijk} &:=& \Xi^{(m,n)}_{ij} e^{(n,m)}_{k},  \\
		\Xi^{(m,n)\, i} &:=& g^{(m) ij} g^{(n) kl} \left( R_{\Xi} \right)^{(m,n)}_{jkl} .
	\end{eqnarray}
\end{subequations}
$\Xi^{(m,n)\, i}$ is an interesting invariant because it represents the components of a vector; it defines a direction on the parameter space that is invariant, and its real and imaginary parts are observables.

Remember that these invariants are defined for two different subspaces, $\mathcal{H}_{n}\neq \mathcal{H}_{m}$. However, notice that it is possible to construct one-subspace invariants, such as $g^{(n)}_{ij}$ or $F^{(n)}_{ij}$, if we consider an additional sum over the subspaces $\mathcal{H}_{m}$, with $m\neq n$. Finally, there exists a real 3-form invariant. But we postpone its definition until the end of the next section. For more properties of the invariants, see appendix~\ref{Ap_prop}.

\section{Formulation with differential forms}
\label{Sec_df}

Before exploring applications of the previous formalism, we summarize the relevant quantities using a simplified notation combined with differential forms. This approach not only facilitates the expressions but also emphasizes the geometric meaning of the objects under consideration.

In this work, we use a notation that explicitly captures all key information defining each geometric object. For instance, the $\mathcal{N}$-bein $e^{(m,n)}_{i\; MN}$ represents the variation of the $i$th parameter of the state $\ket{n_{N}}$ projected onto a different state $\ket{m_{M}}$. While this detailed notation is valuable for tracking the relevant components of each object, it can become cumbersome as we become more familiar with the framework. To address this, we introduce a simplified notation that facilitates manipulation and aligns more closely with conventions in other areas of physics, such as gravitational theories.

For the simplified notation, we define our $\mathcal{N}$-bein as
\begin{equation}
	e_{i}{}_{M}{}^{N} := \mathrm{i} \bk{m_{M}}{\partial_{i} n_{N}}. 
\end{equation}
Notice that we omit the indices labeling the energy levels, with the understanding that the associated uppercase letters are related to such levels. Furthermore, the position of the index refers to its origin; it is a superscript when it is associated with a ket $\ket{\cdot}$ and identifies a column in the matrix representation, whereas a subscript is related to a bra $\bra{\cdot}$ and corresponds to a row in the same representation. Also, notice that the conjugation property now acts as a transposition of the indices, so
\begin{equation}
	e_{i \; M}{}^{N \, \ast} = e_{i \; N}{}^{M}.
\end{equation}
Similarly, the WZ connection is
\begin{equation}
	A_{i \; N_{1}}{}^{N_{2}} := \mathrm{i} \bk{n_{N_{1}}}{\partial_{i} n_{N_{2}}}.
\end{equation}
Here, $A_{i \; N_{1}}{}^{N_{2}}$ only has $N$ indices; then it is understood that the connection is defined with states belonging only to $\mathcal{H}_{n}$. Meanwhile, the QGT for the states on $\mathcal{H}_{n}$ is
\begin{equation}
	Q_{ij \; N_{1}}{}^{N_{2}} = \sum_{m\neq n} e_{i\; M}{}^{N_{1} \, \ast} e_{j\; M}{}^{N_{2}} = \sum_{m\neq n} e_{i\; N_{1}}{}^{M} e_{j\; M}{}^{N_{2}}.
\end{equation}
Another important characteristic of this notation is the sum convention over repeated indices, one up and one down. Also, we show the sum over the states $m$. Even though the state is not shown explicitly, the existence of the $M$ index implies that we are referring to a state $m$. We use this convention throughout this section.

Now that we have clarified the notation, we define the differential forms for the formalism. Given two tensors $\mathcal{T}$ and $\mathcal{T}^{\prime}$ of the type ${r}\choose{s}$ and ${r^{\prime}}\choose{s^{\prime}}$, respectively, the tensor product $\mathcal{T} \otimes \mathcal{T}^{\prime}$ is a ${r+r^{\prime}}\choose{s+s^{\prime}}$-type tensor. Additionally, for two 1-forms $\boldsymbol{\alpha}$ and $\boldsymbol{\beta}$, we define the symmetric and anti-symmetric tensor products
\begin{subequations}
	\begin{eqnarray}
		\boldsymbol{\alpha} \boldsymbol{\beta} &:=& \mbox{Sym}(\boldsymbol{\alpha} \otimes \boldsymbol{\beta}) = \frac{1}{2} ( \boldsymbol{\alpha} \otimes \boldsymbol{\beta} + \boldsymbol{\beta} \otimes \boldsymbol{\alpha} ), \\
		\boldsymbol{\alpha} \wedge \boldsymbol{\beta} &:=& \mbox{ASym}(\boldsymbol{\alpha} \otimes \boldsymbol{\beta}) = \frac{1}{2} ( \boldsymbol{\alpha} \otimes \boldsymbol{\beta} - \boldsymbol{\beta} \otimes \boldsymbol{\alpha} ).
	\end{eqnarray}
\end{subequations}
Here, the symbol ``$\wedge$'' represents the wedge product. In general, it maps a $p$-form and a $q$-form into a $(p+q)$-form. Moreover, we use ``$d$'' for the exterior derivative; it takes a $p$-form into a $(p+1)$-form.

Under the new notation, the 1-form $\mathcal{N}$-bein is
\begin{equation}
	\label{e_form}
	\boldsymbol{e}_{M}{}^{N} := \mathrm{i} \bk{m_{M}}{\partial_{i} n_{N}} d\lambda^{i} = \mathrm{i} \bk{m_{M}}{d n_{N}}. 
\end{equation}
Or, equivalently:
\begin{equation}
	\label{e_for2}
	\boldsymbol{e}_{M}{}^{N} = \mathrm{i} \dfrac{\bkf{m_{M}}{d \hat{H}}{n_{N}}}{E_{n} - E_{m}},
\end{equation}
where $d\hat{H} := \partial_{i} \hat{H} d\lambda^{i}$. From an intuitive standpoint, each term $\boldsymbol{e}_{M}{}^{N}$ is a differential form acting as an entry on a $d_{m}\times d_{n}$ rectangular matrix. Similarly, the WZ 1-form connection  
\begin{equation}
    \boldsymbol{A}_{N_{1}}{}^{N_{2}} = \mathrm{i} \bk{n_{N_{1}}}{dn_{N_{2}}}
\end{equation}
and the QGT 
\begin{equation}
	\boldsymbol{Q}_{N_{1}}{}^{N_{2}} = Q_{ij \, N_{1}}{}^{N_{2}} d\lambda^{i} \otimes d\lambda^{j} =\sum_{m\neq n}  \boldsymbol{e}_{N_{1}}{}^{M} \otimes \boldsymbol{e}_{M}{}^{N_{2}}  
\end{equation}
are the components of a $d_{n}\times d_{n}$ square matrix.

The tensor $\boldsymbol{Q}_{N_{1}}{}^{N_{2}}$ is composed by the metric $\boldsymbol{g}_{N_{1}}{}^{N_{2}}=g_{ij\; N_{1}}{}^{N_{2}} d\lambda^{i} d\lambda^{j}$ and the curvature of the WZ connection $\boldsymbol{F}_{N_{1}}{}^{N_{2}} = (1/2) F_{ij\; N_{1}}{}^{N_{2}} d \lambda^{i} \wedge d \lambda^{j} $, namely
\begin{equation}
	\boldsymbol{Q}_{N_{1}}{}^{N_{2}} = \boldsymbol{g}_{N_{1}}{}^{N_{2}} - \mathrm{i}  \boldsymbol{F}_{N_{1}}{}^{N_{2}}.
\end{equation}
Explicitly, in terms of the $\mathcal{N}$-bein, we have
\begin{eqnarray}
	\boldsymbol{g}_{N_{1}}{}^{N_{2}} &=& \mbox{Sym}\left(\sum_{m\neq n} \boldsymbol{e}_{N_{1}}{}^{M} \otimes \boldsymbol{e}_{M}{}^{N_{2}} \right) =\sum_{m\neq n} \boldsymbol{e}_{N_{1}}{}^{M}  \boldsymbol{e}_{M}{}^{N_{2}} , \\
	\boldsymbol{F}_{N_{1}}{}^{N_{2}} &=&  \mathrm{i} \, \mbox{ASym}\left( \sum_{m\neq n} \boldsymbol{e}_{N_{1}}{}^{M} \otimes \boldsymbol{e}_{M}{}^{N_{2}} \right) = \mathrm{i} \sum_{m\neq n} \boldsymbol{e}_{N_{1}}{}^{M} \wedge \boldsymbol{e}_{M}{}^{N_{2}}.
\end{eqnarray}
Since the WZ connection is non-Abelian, its curvature is
\begin{equation}
	\boldsymbol{F}_{N_{1}}{}^{N_{2}} := d\boldsymbol{A}_{N_{1}}{}^{N_{2}} - \mathrm{i} \boldsymbol{A}_{N_{1}}{}^{N_{3}} \wedge \boldsymbol{A}_{N_{3}}{}^{N_{2}}.
\end{equation}

On the other hand, with this notation, the torsion is
\begin{equation}
	\boldsymbol{T}_{M}{}^{N} := D \boldsymbol{e}_{M}{}^{N} = d \boldsymbol{e}_{M}{}^{N} - \mathrm{i} \boldsymbol{A}_{M}{}^{M_{1}} \wedge \boldsymbol{e}_{M_{1}}{}^{N} + \mathrm{i} \boldsymbol{A}_{N_{1}}{}^{N} \wedge \boldsymbol{e}_{M}{}^{N_{1}}.
\end{equation}
Notice the familiar rule where the lower indices transform with a negative sign, while the upper indices transform with a positive sign. From here, we compute the covariant derivatives and obtain the identities:
\begin{subequations}
	\begin{eqnarray}
		\label{B1}
		D \boldsymbol{F}_{N_{1}}{}^{N_{2}} &=& d\boldsymbol{F}_{N_{1}}{}^{N_{2}} - \mathrm{i} \boldsymbol{A}_{N_{1}}{}^{N_{3}} \wedge \boldsymbol{F}_{N_{3}}{}^{N_{2}} + \mathrm{i} \boldsymbol{A}_{N_{3}}{}^{N_{2}} \wedge \boldsymbol{F}_{N_{1}}{}^{N_{3}}  = 0, \\ 
		\label{B2}
		D \boldsymbol{T}_{M}{}^{N} &=& d\boldsymbol{T}_{M}{}^{N} - \mathrm{i} \boldsymbol{A}_{M}{}^{M_{1}} \wedge \boldsymbol{F}_{M_{1}}{}^{N} + \mathrm{i} \boldsymbol{A}_{N_{1}}{}^{N} \wedge \boldsymbol{T}_{M}{}^{N_{1}} \nonumber \\
        &=& \mathrm{i} \boldsymbol{F}_{N_{1}}{}^{N} \wedge \boldsymbol{e}_{M}{}^{N_{1}} - \mathrm{i} \boldsymbol{F}_{M}{}^{M_{1}} \wedge \boldsymbol{e}_{M_{1}}{}^{N}.
	\end{eqnarray}
\end{subequations}

The torsion is also the anti-symmetric part of the two-state QGT, which is 
\begin{equation}
	\boldsymbol{M}_{M}{}^{N} = \sum_{l\neq n, m}  \boldsymbol{e}_{M}{}^{L} \otimes \boldsymbol{e}_{L}{}^{N}.
\end{equation}
In terms of its symmetric and anti-symmetric parts ($\boldsymbol{\mathcal{G}}^{N}{}_{M}$ and $\boldsymbol{T}^{N}{}_{M}$, respectively), it is
\begin{equation}
	\boldsymbol{M}_{M}{}^{N} = \boldsymbol{\mathcal{G}}_{M}{}^{N}  - \mathrm{i} \boldsymbol{T}_{M}{}^{N},
\end{equation}
where
\begin{eqnarray}
	\boldsymbol{\mathcal{G}}_{M}{}^{N} &=& \mbox{Sym}\left(\sum_{l\neq n, m} \boldsymbol{e}_{M}{}^{L} \otimes \boldsymbol{e}_{L}{}^{N} \right) = \sum_{l\neq n, m}   \boldsymbol{e}_{M}{}^{L} \boldsymbol{e}_{L}{}^{N}, \\
	\boldsymbol{T}_{M}{}^{N} &=&  \mathrm{i} \, \mbox{ASym}\left( \sum_{l\neq n, m} \boldsymbol{e}_{M}{}^{L} \otimes \boldsymbol{e}_{L}{}^{N} \right) = \mathrm{i} \sum_{l\neq n, m}  \boldsymbol{e}_{M}{}^{L} \wedge \boldsymbol{e}_{L}{}^{N}.
\end{eqnarray}
For the quantum systems where we have at least three independent parameters ($\mathcal{N}\ge 3$), we define an invariant 3-form for every subspace $\mathcal{H}_{n}$ 
\begin{equation}
	\label{inv_tau}
	\boldsymbol{\tau}^{(n)} := \sum_{m\neq n} \boldsymbol{T}_{M}{}^{N} \wedge \boldsymbol{e}_{N}{}^{M}=\sum_{m\neq n} \boldsymbol{e}_{M}{}^{N} \wedge \boldsymbol{T}_{N}{}^{M}.
\end{equation}
This invariant belongs to the family of invariants given in \eqref{inv_R}, but it has two additional restrictions: an anti-symmetrization on the parameter indices and the sum over all the states $m\neq n$. These two additional conditions guarantee that $\boldsymbol{\tau}^{(n)}$ is always an invariant real 3-form. Furthermore, for systems with exactly three independent parameters ($\mathcal{N}=3$), $\boldsymbol{\tau}^{(n)}$ is proportional to the volume form of the parameter space. It is worth mentioning that this invariant is not mentioned in~\cite{Romero2409}; however, deriving its expression for the nondegenerate case is straightforward.
\begin{equation}
	\boldsymbol{\tau}^{(n)} := \sum_{m\neq n} \boldsymbol{T}^{(m,n)} \wedge \boldsymbol{e}^{(n,m)}= \sum_{m\neq n} \boldsymbol{e}^{(m,n)} \wedge \boldsymbol{T}^{(n,m)}.
\end{equation}
This 3-form is integrable over a three-dimensional parameter manifold, defining a scalar invariant independent of the parameters, quite similar to a Chern-Simons non-Abelian 3-form \cite{Chern-Simons}, and directly related to the torsional topological invariants of Nieh-Yan \cite{Nieh1,Nieh}, see also \cite{Chandia}. This property implies the emergence of a new topological invariant within the underlying parameter space. This invariant might be highly relevant to describe third-order nonlinear transport responses \cite{Agarwal2}.

\section{Illustrative Example: Coupled harmonic oscillators in a constant electric field }
\label{Sec_ejem}

\subsection{The system and its solution}

To illustrate the advantages and applications of our formalism, we study a quantum system consisting of three identical coupled harmonic oscillators. Additionally, we add a linear term in each direction, typically associated with an electric field. Complementary material for this section is presented in appendix \ref{Ap_Ex_extra}. The Hamiltonian describing this system is
\begin{equation}\label{hamil21}
	\hat{H}=\frac{1}{2} \biggl\{\sum_{a=1}^{3}(\hat{p}_{a}^{2}+k\hat{q}_{a}^{2})+k'\sum_{a<b}^{3}(\hat{q}_{a}-\hat{q}_{b})^{2}\biggr\}+X\sum_{a=1}^{3}\hat{q}_{a} ,
\end{equation}
where $k$ is the oscillators' constant, $k'$ is a coupling parameter, and $X$ represents the field strength. The position and momentum operators for each particle are $\hat{q}_{a}$ and $\hat{p}_{a}$, respectively, with $a=1,2,3$.\\

To solve the system given by \eqref{hamil21}, we need to find the normal modes and uncouple the oscillators. Thus, we write \eqref{hamil21} as 
\begin{equation}\label{hamil2}
	\hat{H}=\frac{1}{2}\left( \hat{\mathbf{p}}^{T}\hat{\mathbf{p}}+\hat{\mathbf{q}}^{T}\mathds{K}\hat{\mathbf{q}}\right)+X\sum_{a=1}^{3}\hat{q}_{a} ,
\end{equation}
where $\hat{\mathbf{q}}^{T}=\left( \hat{q}_{1}\; \hat{q}_{2} \; \hat{q}_{3} \right)$ and $\hat{\mathbf{p}}^{T}=\left( \hat{p}_{1}\; \hat{p}_{2} \; \hat{p}_{3} \right)$. Hence, the problem becomes the diagonalization of the matrix
\begin{equation}\label{kmatrix}
	\mathds{K}=\begin{pmatrix} k+2k' & -k' & -k'\\  -k' & k+2k' & -k' \\
		-k' & -k' & k+2k'
	\end{pmatrix}. 
\end{equation}
The eigenvalues of $\mathds{K}$ are the normal modes' frequencies $\omega_{1}^{2}:=k$ and $\omega_{2}^{2}=\omega_{3}^{2}:=k+3k'$. Notice the degeneracy in the frequencies; it will result in a degeneracy in the energy spectrum. Then, using the diagonalization matrix 
\begin{equation}
	\mathds{U}=\begin{pmatrix} 1/\sqrt{3} & -1/\sqrt{2} & -1/\sqrt{6}\\  1/\sqrt{3} & 0 & 2/\sqrt{6}\\ 1/\sqrt{3} & 1/\sqrt{2} & -1/\sqrt{6} \end{pmatrix},
\end{equation}
the Hamiltonian becomes 
\begin{align}\label{hamilt4}
	\hat{H} &=\frac{1}{2}\left( \hat{\mathbf{P}}^{T}\hat{\mathbf{P}}+\hat{\mathbf{Q}}^{T}\Omega^{2}\hat{\mathbf{Q}}\right)+\sqrt{3} X \hat{Q}_{1} \nonumber\\
	&=\frac{1}{2}\sum_{a=1}^{3}\left( \hat{P}_{a}^{2}+\omega_{a}^{2}\hat{Q}_{a}^{2} \right) + \sqrt{3} X \hat{Q}_{1}.
\end{align}
We defined $\Omega^{2}:=\mathds{U}^{T}\mathds{K}\mathds{U}=\text{diag} (\omega_{1}^{2}, \omega_{2}^{2}, \omega_{3}^{2})=\text{diag}(k, k+3k', k+3k')$ and the normal coordinates are
\begin{eqnarray}\label{coor2}
	\hat{\mathbf{Q}}&=&\mathds{U}^{T}\hat{\mathbf{q}}=\begin{pmatrix} \hat{Q}_{1} \\  \hat{Q}_{2}\\ \hat{Q}_{3} \end{pmatrix}= \begin{pmatrix} \frac{1}{\sqrt{3}}(\hat{q}_{1}+\hat{q}_{2}+\hat{q}_{3}) \\  \frac{1}{\sqrt{2}}(-\hat{q}_{1}+\hat{q}_{3})\\ \frac{1}{\sqrt{6}}(-\hat{q}_{1}+2\hat{q}_{2}-\hat{q}_{3}) \end{pmatrix},\\
	\label{coor3}
	\hat{\mathbf{P}}&=&\mathds{U}^{T}\hat{\mathbf{p}}=\begin{pmatrix} \hat{P}_{1} \\  \hat{P}_{2} \\ \hat{P}_{3}\end{pmatrix}=\begin{pmatrix} \frac{1}{\sqrt{3}}(\hat{p}_{1}+\hat{p}_{2}+\hat{p}_{3}) \\  \frac{1}{\sqrt{2}}(-\hat{p}_{1}+\hat{p}_{3})\\ \frac{1}{\sqrt{6}}(-\hat{p}_{1}+2\hat{p}_{2}-\hat{p}_{3}) \end{pmatrix}.
\end{eqnarray}

The Hamiltonian (\ref{hamilt4}) is uncoupled and is the sum of three harmonic oscillators. The normal coordinates are such that one mode is aligned with the field of strength $X$, whereas the remaining modes are just simple oscillators orthogonal to that field. Therefore, the solution of the system is straightforward. We solve each oscillator separately, so the complete solution is the product of the three individual oscillators:
\begin{eqnarray}\label{wavefunction}
	\Psi_{n_{1},n_{2},n_{3}}(Q_{1},Q_{2},Q_{3};X,k,k') &=& \left(\frac{\omega_{1}\omega_{2}^{2}}{\hbar^{3}}\right)^{1/4}\chi_{n_{1}}\left[\left(Q_{1}+\frac{\sqrt{3} X}{\omega_{1}^{2}}\right)\sqrt{\frac{\omega_{1}}{\hbar}}\right]\chi_{n_{2}}\left[ Q_{2}\sqrt{\frac{\omega_{2}}{\hbar}}\right] \nonumber \\
    &&\times \chi_{n_{3}}\left[ Q_{3}\sqrt{\frac{\omega_{2}}{\hbar}}\right].
\end{eqnarray}
Here, $n_{1}, n_{2}, n_{3}=0,1,2,\dots$, and we introduced 
\begin{equation}
	\chi_{n_{a}}(\xi) := \left(\frac{1}{2^{n_{a}}n_{a}!\sqrt{\pi}}\right)^{1/2}e^{-\xi^{2}/2}H_{n_{a}}(\xi),
\end{equation}
where $H_{n_{a}}(\xi)$ is the Hermite polynomial of degree $n_{a}$. Furthermore, the functions $\chi_{n_{a}}(\xi)$ satisfy
\begin{eqnarray}
	&\displaystyle{\int} d \xi \chi_{n_{a}}(\xi) \chi_{m_{b}}(\xi) = \delta_{n_{a}m_{b}}, \\
	&\dfrac{d \chi_{n_{a}}(\xi)}{d \xi} = \sqrt{\dfrac{n_{a}}{2}} \chi_{n_{a}-1} (\xi) - \sqrt{\dfrac{n_{a}+1}{2}} \chi_{n_{a}+1} (\xi), \\
	&\xi \chi_{n_{a}}(\xi) = \sqrt{\dfrac{n_{a}}{2}} \chi_{n_{a}-1} (\xi) + \sqrt{\dfrac{n_{a}+1}{2}} \chi_{n_{a}+1} (\xi).
\end{eqnarray}
On the other hand, the energy of the system depends on the three parameters, $X$, $k$, and $k^{\prime}$. The dependency on the latter two is through the frequencies $\omega_{1}$ and $\omega_{2}$; meanwhile, the parameter $X$ appears as a constant energy shift, explicitly:
\begin{equation}
	E_{n_{1},n_{2},n_{3}}(X,k,k')=\left( n_{1}+\frac{1}{2}\right)\hbar\omega_{1}+\left( n_{2}+n_{3}+1\right)\hbar\omega_{2}-\frac{3X^{2}}{2\omega_{1}^{2}}.
\end{equation}

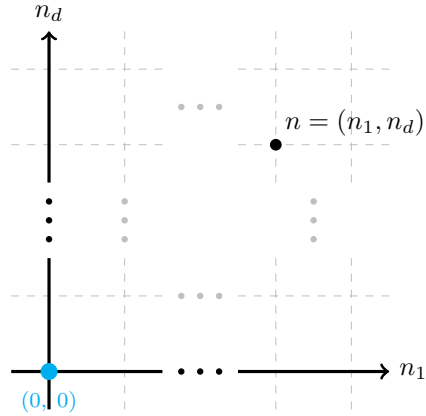
\begin{figure}[h!]
	\centering
	\begin{tikzpicture}
		
		\draw[very thin, color=gray!50, dashed] (-0.5,-0.5) grid (1.5,1.5);
		\draw[very thin, color=gray!50, dashed] (-0.5,2.5) grid (1.5,4.5);
		\draw[very thin, color=gray!50, dashed] (2.5,-0.5) grid (4.5,1.5);
		\draw[very thin, color=gray!50, dashed] (2.5,2.5) grid (4.5,4.5);
		
		\filldraw[color=gray!50] (2,1) circle (1pt);
		\filldraw[color=gray!50] (1.75,1) circle (1pt);
		\filldraw[color=gray!50] (2.25,1) circle (1pt);
		
		\filldraw[color=gray!50] (2,3.5) circle (1pt);
		\filldraw[color=gray!50] (1.75,3.5) circle (1pt);
		\filldraw[color=gray!50] (2.25,3.5) circle (1pt);

		\filldraw[color=gray!50] (1,2) circle (1pt);
		\filldraw[color=gray!50] (1,1.75) circle (1pt);
		\filldraw[color=gray!50] (1,2.25) circle (1pt);
		
		\filldraw[color=gray!50] (3.5,2) circle (1pt);
		\filldraw[color=gray!50] (3.5,1.75) circle (1pt);
		\filldraw[color=gray!50] (3.5,2.25) circle (1pt);

		\draw[-, very thick] (-0.5,0) -- (1.5,0);
		\filldraw (2,0) circle (1pt);
		\filldraw (1.75,0) circle (1pt);
		\filldraw (2.25,0) circle (1pt);
		\draw[->, very thick] (2.5,0) -- (4.5,0) node[right] {$n_{1}$};
		
		\draw[-, very thick] (0,-0.5) -- (0,1.5);
		\filldraw (0,2) circle (1pt);
		\filldraw (0,1.75) circle (1pt);
		\filldraw (0,2.25) circle (1pt);
		\draw[->, very thick] (0,2.5) -- (0,4.5) node[above] {$n_{d}$};
		
		\filldraw[cyan] (0,0) circle (3pt) node[below=4pt] {\footnotesize $(0,\; 0)$};
		\filldraw[black] (3,3) circle (2pt) node[above right] {$n=(n_{1},n_{d})$};
	\end{tikzpicture}
	\caption{Diagram representing an arbitrary state $n=(n_{1}, n_{d})$ with energy $E_{n}$ (black), and the ground state being at coordinate origin (blue).}
	\label{Fig:states}
\end{figure}

Since the sum of two natural numbers ($n_{2} + n_{3}$) can be expressed in ($n_{2} + n_{3} + 1$)-different ways, all the states 
\begin{equation}
	\label{level_n}
	(n_{1}, n_{2}+n_{3},0),\,(n_{1}, n_{2}+n_{3}-1,1),\, \dots ,\, (n_{1}, n_{2}, n_{3}) ,\, \dots ,\, (n_{1}, 1, n_{2}+n_{3}-1),\, (n_{1}, 0, n_{2}+n_{3})
\end{equation}
share the same energy. Therefore, we identify the state $n$ with the doublet $n=(n_{1}, n_{d}:= n_{2} + n_{3})$, which has degeneracy $d_{n} := n_{d}+1$. Thus, the energy for each level is 
\begin{equation}
	\label{En}
	E_{n}:= E_{(n_{1},n_{d})}=\left( n_{1}+\frac{1}{2}\right)\hbar\omega_{1}+\left( n_{d}+1\right)\hbar\omega_{2}-\frac{3X^{2}}{2\omega_{1}^{2}}.
\end{equation}
Using this notation, we represent each state with an energy $E_{n}$ as a point in a grid diagram with coordinates $(n_{1},n_{d})$, see figure~\ref{Fig:states}, where the ground state is at the origin at the point $(0,0)$. Since the degeneracy $d_{n}$ on a given energy level depends only on the quantum number $n_{d}$, in the diagram, the higher the degeneracy of a state, the higher its position $n$ the $n_{d}$ axis will be. Furthermore, all the states with the same degeneracy are found along the same horizontal line, with the non-degenerate states being along the $n_{d}=0$ axis.	

To be consistent with the notation of the previous sections, we use the letter $N$ to label all states with the same energy. From the original triplet of quantum numbers $(n_{1}, n_{2}, n_{3})$, we define 
\begin{equation}
	N := n_{3} + 1. 
\end{equation}
Therefore, all the states in \eqref{level_n} correspond to the same energy level $n$ with the label $N$ that goes from 1 to $d_{n}$ from left to right. Notice that the map $(n_{1}, n_{2}, n_{3}) \mapsto (n,N):= ((n_{1}, n_{2} + n_{3}), n_{3}+1)$ is a bijection with its inverse given by $(n_{1}, n_{2}, n_{3})=(\pi_{1}(n), \pi_{2}(n)-N+1,N-1)$. Here $\pi_{1}(n)=n_{1}$ and $\pi_{2}(n)=n_{d}$ give the first and second entry of the doublet $n=(n_{1},n_{d})$, respectively.\\

\subsection{$\mathcal{N}$-bein computation}

\begin{figure}[h!]
	\centering
	\begin{tikzpicture}[scale=0.8]
		
		\draw[very thin, color=gray!50, dashed] (-3.5,-3.5) grid (3.5,3.5);
		
		\draw[<->, very thick] (-3.5,0) -- (3.5,0) node[right] {$n_{1}$};
		\draw[<->, very thick] (0,-3.5) -- (0,3.5) node[above] {$n_{d}$};
		
		\filldraw[cyan] (0,0) circle (4pt) node[below=4pt] {\footnotesize ${(n_{1},\; n_{d})}$};
		\filldraw[black] (1,0)  circle (3pt) node[above] {\tiny ${(n_{1}+1,n_{d})}$};
		\filldraw[black] (-1,0) circle (3pt)node[above] {\tiny ${(n_{1}-1,n_{d})}$};
		\filldraw[black] (2,0) circle (3pt) node[below=4pt] {\tiny ${(n_{1}+2,n_{d})}$};
		\filldraw[black] (-2,0)  circle (3pt) node[below=4pt] {\tiny ${(n_{1}-2,n_{d})}$};
		\filldraw[black] (0,2) circle (3pt) node[above right] {\tiny ${(n_{1},n_{d}+2)}$};
		\filldraw[black] (0,-2) circle (3pt) node[below right] {\tiny ${(n_{1},n_{d}-2)}$};
	\end{tikzpicture}
	\caption{Diagram showing all the states (black dots) related through a parameter variation with the initial state $n=(n_{1}, n_{d})$  (blue dot). They represent all the non-zero $\mathcal{N}$-beins.}
	\label{Fig:Nbein_ex}
\end{figure}
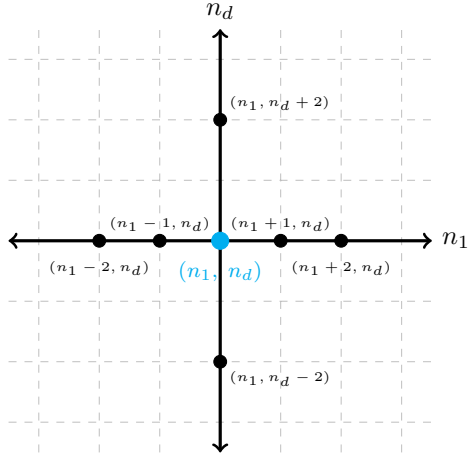

Now that we have established the notation, we compute the $\mathcal{N}$-bein $e^{(m,n)}_{i\; MN}$ defined in  \eqref{e_def}. Here, the parameter index $i$ takes the values  $\{ X,\, k,\, k^{\prime}\}$. Thus, given the solution \eqref{wavefunction}, we derive the $\mathcal{N}$-beins from
\begin{equation}
	e^{(m,n)}_{i\; MN} = \mathrm{i} \int\, dQ_{1}dQ_{2}dQ_{3} \Psi^{*}_{m_{1},m_{2},m_{3}}\partial_{i}\Psi_{n_{1},n_{2},n_{3}}.
\end{equation}
We write the right-hand side of the equation in terms of the original quantum numbers because it is easier to compute the inner product in this way. Nonetheless, once we solve the integral, we write the $\mathcal{N}$-bein using the energy level $n=(n_{1}, n_{d})$ and its position $N$ within the degenerate subspace, explicitly:
\begin{eqnarray}
	\label{N-bein_ex}
	e^{(m,n)}_{i\; MN} &=& \mathrm{i} \Bigg\lbrace \sqrt{\dfrac{3 \omega_{1}}{2 \hbar}}   \left[ \sqrt{n_{1}} \delta_{m_{1},n_{1}-1} - \sqrt{n_{1}+1} \delta_{m_{1},n_{1}+1} \right] \partial_{i}\left(\frac{X}{k}\right)  \nonumber \\
	&&+ \frac{1}{4} \left[ \sqrt{n_{1}(n_{1}-1)} \delta_{m_{1},n_{1}-2} - \sqrt{(n_{1}+1) (n_{1}+2)} \delta_{m_{1},n_{1}+2} \right] \partial_{i} \ln \omega_{1}\Bigg\rbrace \delta_{m_{d},n_{d}} \delta_{M,N} \nonumber \\
	&&+\frac{\mathrm{i}}{4}\delta_{m_{1},n_{1}} \bigg\lbrace \delta_{m_{d},n_{d}-2} \left[ \sqrt{(n_{d}-N)(n_{d}-N+1)} \delta_{M,N} + \sqrt{(N-1)(N-2)} \delta_{M,N-2} \right] \nonumber\\
	&& - \delta_{m_{d},n_{d}+2} \left[ \sqrt{(n_{d}-N+2)(n_{d}-N+3)} \delta_{M,N} + \sqrt{N(N+1)}\delta_{M, N+2} \right]  \Bigg\rbrace \partial_{i} \ln \omega_{2}. 
\end{eqnarray}
Recall that the $\mathcal{N}$-bein represents a transition from an initial state $\ket{n_{N}}$ to another $\ket{m_{M}}$ after a perturbation in the $i$th parameter. Hence, since $n=(n_{1}, n_{d})$ labels the energy level, from the last equation we read six possible energy-level changes after a parameter variation. They are from $(n_{1},n_{d})$ to $(n_{1} \pm 1, n_{d})$, $(n_{1} \pm 2, n_{d})$, or $(n_{1}, n_{d} \pm 2)$. Schematically, we represent them in a diagram, see figure~\ref{Fig:Nbein_ex}. In such a figure, the blue dot at $(n_{1}, n_{d})$ is our initial state, and the black dots indicate the energy levels with a non-zero $\mathcal{N}$-bein, i.e., they are the reachable states after one parameter variation.

We divide the energy-level changes discussed above into three cases. The first case occurs through a variation on the $X$ or $k$ parameter, and it results in a change from $(n_{1}, n_{d})$ to $(n_{1} \pm 1, n_{d})$. Namely, the $\mathcal{N}$-beins are 
\begin{subequations}
	\begin{eqnarray}
		\label{e1m}
		e^{((n_{1}-1,n_{d}), \,(n_{1},n_{d}))}_{i\; M, N} &=& \mathrm{i} \sqrt{\frac{3n_{1}\omega_{1}}{2\hbar}} \left(\partial_{i}\frac{X}{k}\right) \delta_{M, N}, \\
		\label{e1p}
		e^{((n_{1}+1,n_{d}),\, (n_{1},n_{d}))}_{i\; M N} &=& - \mathrm{i} \sqrt{\frac{3(n_{1}+1)\omega_{1}}{2\hbar}} \left( \partial_{i} \frac{X}{k}\right) \delta_{M, N}.
	\end{eqnarray}
\end{subequations}
Similarly, when we vary the parameter $k$, we have a change to $n_{1} \pm 2 $. The $\mathcal{N}$-beins for this case are
\begin{subequations}
	\begin{eqnarray}
		\label{e2m}
		e^{((n_{1}-2,n_{d}),\, (n_{1},n_{d}))}_{i\; M N} &=& \frac{\mathrm{i}}{4} \sqrt{n_{1}(n_{1}-1)}  \left( \partial_{i} \ln \omega_{1} \right) \delta_{M, N} ,\\
		\label{e2p}
		e^{((n_{1}+2,n_{d}),\, (n_{1},n_{d}))}_{i\; M N} &=& - \frac{\mathrm{i}}{4} \sqrt{(n_{1}+1)(n_{1}+2)}  \left( \partial_{i} \ln \omega_{1} \right)\delta_{M,  N}.
	\end{eqnarray}
\end{subequations}
Both of these cases only change the number $n_{1}$ and are represented by black dots along the $n_{1}$ axis in figure~\ref{Fig:Nbein_ex}. They relate the original state $n=(n_{1}, n_{d})$ to another with the same degeneracy, and the factor $\delta_{M, N}$ indicates that we are maintaining the same direction $N$, even though we change from one subspace $\mathcal{H}_{n}$ to another. In contrast, the third case occurs through a variation on $k$ or $k^{\prime}$, and it corresponds to a change from $(n_{1}, n_{d})$ to $(n_{1}, n_{d} \pm 2)$. Here, the $\mathcal{N}$-beins are 
\begin{subequations}
	\begin{eqnarray}
		\label{e3m}
		e^{((n_{1},n_{d}-2),\, (n_{1},n_{d}))}_{i\; MN} &=& \frac{\mathrm{i}}{4} \left( \partial_{i} \ln \omega_{2} \right) \Big[\sqrt{(n_{d}-N)(n_{d}-N+1)}\delta_{M,N} \nonumber\\
            && + \sqrt{(N-1)(N-2)}\delta_{M,N-2} \Big] ,\\
		\label{e3p}
		e^{((n_{1},n_{d}+2),\, (n_{1},n_{d}))}_{i\; MN} &=& - \frac{\mathrm{i}}{4}  \left( \partial_{i} \ln \omega_{2} \right) \Big[ \sqrt{(n_{d}-N+2)(n_{d}-N+3)} \delta_{M,N} \nonumber\\ 
            && + \sqrt{N(N+1)}\delta_{M,N+2} \Big]. 
	\end{eqnarray}
\end{subequations}
Hence, it is possible to change from one state to another with different degeneracy, so the degenerate part of the $\mathcal{N}$-beins is a rectangular matrix of dimension $d_{m}\times d_{n}$, with $d_{m} = d_{n} \mp 2$, respectively. Moreover, besides the change of energy level, we also identify a possible change in $N$, which represents a direction in the subspace $\mathcal{H}_{n}$. The states related through these $\mathcal{N}$-beins are represented in the diagram of figure~\ref{Fig:Nbein_ex} by the black dots along the $n_{d}$-axis. From the three cases, only a variation on $k$ could result in any of the three possible scenarios, whereas a perturbation on $X$ or $k^{\prime}$ leads to the first case or to the third case, exclusively. 

The study of $\mathcal{N}$-beins not only exposes correlations due to parameter variations but also determines the impossibility to reach certain states. For instance, a variation in the field strength $X$ will not change the state's degeneracy. Furthermore, from figure~\ref{Fig:Nbein_ex}, we see that it is not possible to reach a neighboring state whose degeneracy changes by one unit; this effect is more pronounced when considering two parameter variations, as shown in the next subsection.

Interestingly, notice that the $\mathcal{N}$-beins possess a factorizable structure composed of two parts, one depending entirely on the parameters and one encoding the information about the degeneracy. Thus, we use matrix notation to rewrite the $\mathcal{N}$-beins in a simplified manner. For this convention, when we write any quantity with indices between two square brackets, we are referring to the matrix whose components are given by such a quantity,  e.g., $\mfs{\delta_{N_{1}, N_{2}}} = \mathds{1}_{d_{n}}$ is the $d_{n}\times d_{n}$ identity matrix. Hence, we write the six non-zero $\mathcal{N}$-beins as:
\begin{subequations}
	\begin{eqnarray}
		\label{e1m_M}
		e^{((n_{1}-1,n_{d}),\, (n_{1},n_{d}))} &:=&\mfs{e^{((n_{1}-1,n_{d}),\, (n_{1},n_{d}))}_{i\; N_{1} N_{2}}} = \mathrm{i} \sqrt{n_{1}} \mathbf{E}_{1} \Kp \mathds{1}_{d_{n}} \\
		e^{((n_{1}+1,n_{d}),\, (n_{1},n_{d}))} &:=& \mfs{e^{((n_{1}+1,n_{d}),\, (n_{1},n_{d}))}_{i\; N_{1} N_{2}}} = - \mathrm{i} \sqrt{n_{1}+1} \mathbf{E}_{1} \Kp \mathds{1}_{d_{n}} \\
		&& \nonumber \\                            
		e^{((n_{1}-2,n_{d}),\, (n_{1},n_{d}))} &:=& \mfs{e^{((n_{1}-2,n_{d}),\, (n_{1},n_{d}))}_{i\; N_{1} N_{2}}}  = \mathrm{i} \sqrt{n_{1}(n_{1}-1)}  \mathbf{E}_{2} \Kp \mathds{1}_{d_{n} },\\
		e^{((n_{1}+2,n_{d}),\, (n_{1},n_{d}))} &:=& \mfs{e^{((n_{1}+2,n_{d}),\, (n_{1},n_{d}))}_{i\; N_{1} N_{2}}}  = - \mathrm{i} \sqrt{(n_{1}+1)(n_{1}+2)}  \mathbf{E}_{2} \Kp \mathds{1}_{d_{n}}, \\
		&& \nonumber \\                            
		e^{((n_{1},n_{d}-2),\, (n_{1},n_{d}))} &:=& \mfs{e^{((n_{1},n_{d}-2),\, (n_{1},n_{d}))}_{i\; NM} } = \mathrm{i} \mathbf{E}_{3} \Kp \mathds{D}(n_{d}),\\
		\label{e3p_M}                              
		e^{((n_{1},n_{d}+2),\, (n_{1},n_{d}))} &:=& \mfs{e^{((n_{1},n_{d}+2),\, (n_{1},n_{d}))}_{i\; NM} } = - \mathrm{i} \mathbf{E}_{3} \Kp \mathds{D}^{T}(n_{d}+2).
	\end{eqnarray}
\end{subequations}
The parameter part is encoded in the row vectors 
\begin{align}
	\mathbf{E}_{1} &:= \sqrt{\dfrac{3}{2 \hbar \omega_{1}^{7}}} \begin{pmatrix}
		\omega_{1}^{2} & -X &0 
	\end{pmatrix}, \\
	\mathbf{E}_{2} &:= \dfrac{1}{8 \omega_{1}^{2}} \begin{pmatrix}
		0 & 1 & 0
	\end{pmatrix}, \\
	\qquad \qquad
	\mathbf{E}_{3} &:= \dfrac{1}{8 \omega_{2}^{2}}  \begin{pmatrix}
		0 & 1 & 3
	\end{pmatrix}, 
\end{align}
obtained by simplifying the derivatives in~\eqref{e1m}--\eqref{e3p}. On the other hand, the symbol $\Kp$ denotes the Kronecker product, indicating that each entry on the vectors is multiplied by either the identity matrix $\mathds{1}_{d_{n}}$ or the rectangular matrices $\mathds{D}(n_{d})$ and $\mathds{D}^{T}(n_{d}+2)$, derived from the part between square brackets in~\eqref{e3m} and \eqref{e3p}, respectively. Explicitly, the components of $\mathds{D}(n_{d})$ are 
\begin{equation}
	D^{(n_{d})}_{MN} := \sqrt{(n_{d}-N)(n_{d}-N+1)}\delta_{M,N} + \sqrt{(N-1)(N-2)}\delta_{M,N-2},
\end{equation}
or in matrix form:
{\small
\begin{equation}
	\mathds{D}(n_{d}) := \begin{pmatrix}
		\sqrt{n_{d}(n_{d}-1)} & 0 &  \sqrt{2}& & \cdots  &   & 0 & 0  & 0\\
		0 & \sqrt{(n_{d} - 1) (n_{d}-2)} &  0 & &\cdots  &   & 0 & 0  & 0\\
		&  & \ddots &   & \ddots &   \\
		\vdots & \vdots &   &  & \vdots & & & \vdots & \vdots \\
		&   & &  & \ddots & & \ddots  \\
		0 & 0 & 0 & & \cdots &  & 0& \sqrt{(n_{d} - 1) (n_{d}-2)}  & 0 \\
		0 & 0 & 0 & & \cdots &  & \sqrt{2} & 0 & \sqrt{n_{d}(n_{d}-1)}  
	\end{pmatrix}.
\end{equation}}
$\mathds{D}(n_{d})$ is a $(d_{n}-2)\times d_{n}$ rectangular matrix, while $\mathds{D}^{T}(n_{d})$ is its transpose. As noted, $\mathds{D}(n_{d})$ is a rectangular matrix because the $\mathcal{N}$-beins relate two states with different degeneracy. Moreover, the non-diagonal structure of these matrices corresponds to a possible change in the direction $N$, the direction within the subspace $\mathcal{H}_{n}$.

The $\mathcal{N}$-beins relate an initial state $n=(n_{1}, n_{d})$ with a different state $m=(m_{1}, m_{d})$. The new state $m$ is offset with respect to $n$ by one or two units in the $n_{1}$ direction or two units in the $n_{d}$ direction. The $\mathcal{N}$-beins associated with changes in $n_{1}$ do not depend on explicitly $n_{d}$ (only on the dimension of $\mathds{1}_{d_{n}}$). Similarly, when the degeneracy number $n_{d}$ changes, the $\mathcal{N}$-bein does not depend on $n_{1}$. This characteristic will be inherited by the subsequent tensors and invariants.\\

\subsection{Calculation of the non-abelian Quantum Geometric Tensor}

The $\mathcal{N}$-beins in the matrix notation facilitate the computation of the QGT and all the other geometric objects. Using~\eqref{qgt_def} and \eqref{e1m_M}--\eqref{e3p_M}, the QGT is:
\begin{eqnarray}
	\label{Q_Matrix}
	\mathds{Q}^{(n)} &:=&  \mfs{Q^{(n)}_{ij\, N_{1} N_{2}}} = \sum_{m\neq n} \mfs{\sum_{M=1}^{d_{m}}  e^{(m,n)\,\ast}_{i\; MN} e^{(m,n)}_{j\; MN}}  = \sum_{m\neq n} \left( e^{(m,n)} \right)^{\dagger} \cdot  e^{(m, n)}   \nonumber \\
	&=& n_{1} \left( \mathbf{E}_{1}^{T} \cdot \mathbf{E}_{1} \right) \Kp\left(  \mathds{1}_{d_{n}} \cdot \mathds{1}_{d_{n}} \right) + (n_{1}+1) \left( \mathbf{E}_{1}^{T} \cdot \mathbf{E}_{1} \right) \Kp\left(  \mathds{1}_{d_{n}} \cdot \mathds{1}_{d_{n}} \right) \nonumber \\
	&&  + n_{1}(n_{1}-1) \left( \mathbf{E}_{2}^{T} \cdot \mathbf{E}_{2} \right) \Kp\left(  \mathds{1}_{d_{n}} \cdot \mathds{1}_{d_{n}} \right) + (n_{1} + 1) (n_{1} + 2) \left( \mathbf{E}_{2}^{T} \cdot \mathbf{E}_{2} \right) \Kp\left(  \mathds{1}_{d_{n}} \cdot \mathds{1}_{d_{n}} \right) \nonumber \\
	&& +  \left( \mathbf{E}_{3}^{T} \cdot \mathbf{E}_{3} \right) \Kp\left(  \mathds{D}^{T}(n_{d}) \cdot \mathds{D}(n_{d}) \right) + \left( \mathbf{E}_{3} \cdot \mathbf{E}_{3}^{T} \right) \Kp\left(  \mathds{D}(n_{d}+2) \cdot \mathds{D}^{T}(n_{d}+2) \right) \nonumber \\
	&=&  (2 n_{1} + 1) \left( \mathbf{E}_{1}^{T} \cdot \mathbf{E}_{1} \right) \Kp \mathds{1}_{d_{n}} + 2 (n_{1}^{2} + n_{1} + 1) \left( \mathbf{E}_{2}^{T} \cdot \mathbf{E}_{2} \right) \Kp \mathds{1}_{d_{n}} \nonumber \\
	&& +  \left( \mathbf{E}_{3}^{T} \cdot \mathbf{E}_{3} \right) \Kp \left[ \mathds{D}^{T}(n_{d}) \cdot \mathds{D}(n_{d}) + \mathds{D}(n_{d}+2) \cdot \mathds{D}^{T}(n_{d} + 2) \right],
\end{eqnarray}
where the dot ``$\cdot$'' is the usual matrix multiplication. From the first to the second line, we expanded the sum accounting only for the non-zero $\mathcal{N}$-beins and used the mix-product property of the Kronecker product:
\begin{equation}
	\left( A \Kp B \right) \cdot \left( C \Kp D \right) = \left( A \cdot C \right) \Kp \left( B \cdot D \right).
\end{equation}
Meanwhile, from the second to the third equality, we just simplified the terms. The computation of the remaining matrix products results in
\begin{eqnarray}
	\mathds{Q}^{(n)} & = & \frac{3(2n_{1} + 1)}{2 \hbar \omega_{1}^{7}}\begin{pmatrix}
		\omega_{1}^{4} & - \omega_{1}^{2} X & 0 \\
		- \omega_{1}^{2} X & X^{2} & 0 \\
		0 & 0 & 0  
	\end{pmatrix} \Kp \mathds{1}_{d_{n}}
	+ \frac{n_{1}^{2} + n_{1} + 1}{32 \omega_{1}^{4}}\begin{pmatrix}
		0 & 0 & 0 \\
		0 & 1 & 0 \\
		0 & 0 & 0  
	\end{pmatrix} \Kp \mathds{1}_{d_{n}} \nonumber \\
 && + \frac{1}{32 \omega_{2}^{4}}\begin{pmatrix}
		0 & 0 & 0 \\
		0 & 1 & 3 \\
		0 & 3 & 9  
	\end{pmatrix} \Kp \mathds{B}(n_{d}),
\end{eqnarray}
where
\begin{equation}
	\mathds{B}(n_{d}) := \dfrac{1}{2} \left[ \mathds{D}^{T}(n_{d}) \cdot \mathds{D}(n_{d}) + \mathds{D}(n_{d}+2) \cdot \mathds{D}^{T}(n_{d} + 2) \right]
\end{equation}
is a square symmetric matrix of dimension $d_{n}\times d_{n}$. Explicitly, its components are
\begin{eqnarray}
	\label{B_Ap}
	B^{(n_{d})}_{N_{1} N_{2}} &:=&  \left[ (n_{d}- N_{1}+1) (n_{d} - N_{1} + 2) + N_{1} ( N_{1} - 1) + 2 \right] \delta_{N_{1}, N_{2}} \nonumber \\
	&& +  \sqrt{N_{1}(N_{1} + 1)(n_{d} - N_{1}) (n_{d}-N_{1}+1)} \delta_{N_{1}, N_{2}-2} \nonumber \\
	&& +  \sqrt{N_{2}(N_{2} + 1)(n_{d} - N_{2}) (n_{d}-N_{2}+1)} \delta_{N_{1}-2, N_{2}}.
\end{eqnarray}

Notice that the components of the QGT are real, so the curvature of the WZ connection is zero. Furthermore, the structure of $\mathds{Q}$ and how we compute it tell us about its composition. The first two terms correspond to state jumps given by $(n_{1}, n_{d}) \rightarrow (n_{1}\pm 1, n_{d})$ and $(n_{1}, n_{d}) \rightarrow (n_{1}\pm 2, n_{d})$, respectively. These possible state changes do not alter the $N$ direction; thus, the part encoding the degeneracy is just the identity matrix. On the other hand, the last part of $\mathds{Q}$ accounts for a degeneracy jump, and the non-diagonal structure of $\mathds{B}(n_{d})$ allows a change in the $N$ direction. \\

\subsection{Two-state QGT}

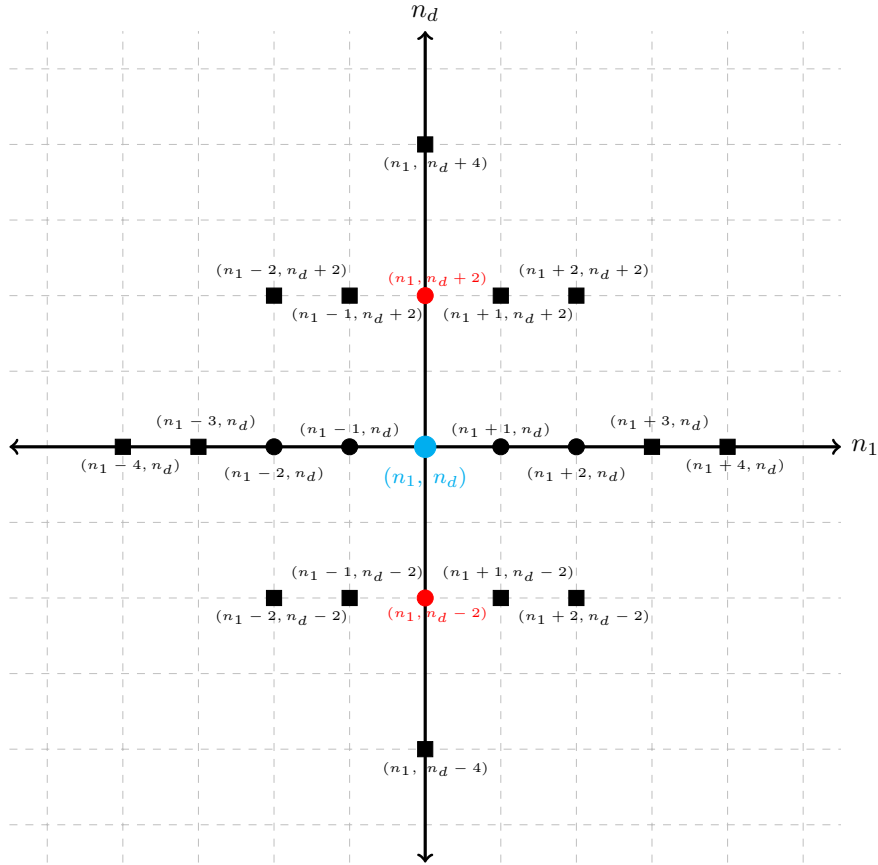
\begin{figure}[h!]
	\centering
	\begin{tikzpicture}[scale=1]
		
		\draw[very thin, color=gray!50, dashed] (-5.5,-5.5) grid (5.5,5.5);
		
		\draw[<->, very thick] (-5.5,0) -- (5.5,0) node[right] {$n_{1}$};
		\draw[<->, very thick] (0,-5.5) -- (0,5.5) node[above] {$n_{d}$};
		
		\filldraw[cyan] (0,0) circle (4pt) node[below=4pt] {\footnotesize ${(n_{1},\; n_{d})}$};
		\filldraw[black] (1,0)  circle (3pt) node[above] {\tiny ${(n_{1}+1,n_{d})}$};
		\filldraw[black] (-1,0) circle (3pt)node[above] {\tiny ${(n_{1}-1,n_{d})}$};
		\filldraw[black] (2,0) circle (3pt) node[below=4pt] {\tiny ${(n_{1}+2,n_{d})}$};
		\filldraw[black] (-2,0)  circle (3pt) node[below=4pt] {\tiny ${(n_{1}-2,n_{d})}$};
		\filldraw[red] (0,2) circle (3pt) node[above] {\tiny $\phantom{....}{(n_{1},n_{d}+2)}$};
		\filldraw[red] (0,-2) circle (3pt) node[below] {\tiny $\phantom{....}{(n_{1},n_{d}-2)}$};
		
		\filldraw[black] (3,0) ++ (-0.1,-0.1) rectangle ++ (0.2, 0.2) node[above] {\tiny ${(n_{1}+3,n_{d})}$};
		\filldraw[black] (-3,0) ++ (-0.1,-0.1) rectangle ++ (0.2, 0.2) node[above] {\tiny ${(n_{1}-3,n_{d})}$};
		\filldraw[black] (4,0) ++ (-0.1,-0.1) rectangle ++ (0.2, 0.2) node[below=4pt] {\tiny ${(n_{1}+4,n_{d})}$};
		\filldraw[black] (-4,0) ++ (-0.1,-0.1) rectangle ++ (0.2, 0.2) node[below=4pt] {\tiny ${(n_{1}-4,n_{d})}$};
		
		\filldraw[black] (1,2) ++ (-0.1,-0.1) rectangle ++ (0.2, 0.2) node[below=4pt] {\tiny ${(n_{1}+1,n_{d}+2)}$};
		\filldraw[black] (-1,2) ++ (-0.1,-0.1) rectangle ++ (0.2, 0.2) node[below=4pt] {\tiny ${(n_{1}-1,n_{d}+2)}$};
		\filldraw[black] (2,2) ++ (-0.1,-0.1) rectangle ++ (0.2, 0.2) node[above] {\tiny ${(n_{1}+2,n_{d}+2)}$};
		\filldraw[black] (-2,2) ++ (-0.1,-0.1) rectangle ++ (0.2, 0.2) node[above] {\tiny ${(n_{1}-2,n_{d}+2)}$};
		
		\filldraw[black] (1,-2) ++ (-0.1,-0.1) rectangle ++ (0.2, 0.2) node[above] {\tiny ${(n_{1}+1,n_{d}-2)}$};
		\filldraw[black] (-1,-2) ++ (-0.1,-0.1) rectangle ++ (0.2, 0.2) node[above] {\tiny ${(n_{1}-1,n_{d}-2)}$};
		\filldraw[black] (2,-2) ++ (-0.1,-0.1) rectangle ++ (0.2, 0.2) node[below=4pt] {\tiny ${(n_{1}+2,n_{d}-2)}$};
		\filldraw[black] (-2,-2) ++ (-0.1,-0.1) rectangle ++ (0.2, 0.2) node[below=4pt] {\tiny ${(n_{1}-2,n_{d}-2)}$};
		
		\filldraw[black] (0,4) ++ (-0.1,-0.1) rectangle ++ (0.2, 0.2) node[below=4pt] {\tiny ${\; (n_{1}, \; n_{d}+4)}$};
		\filldraw[black] (0,-4) ++ (-0.1,-0.1) rectangle ++ (0.2, 0.2) node[below=4pt] {\tiny ${\;(n_{1},\; n_{d}-4)}$};
	\end{tikzpicture}
	\caption{Starting at the blue circle in the state with $n=(n_{1}, n_{d})$, we represent all the energy levels related to it by two parameter variations (black squares and black circles). Additionally, the points with circles (either black or red) are also correlated through one parameter variation ($\mathcal{N}$-beins). The red circles indicate states available through one parameter variation only.}
	\label{Fig:Mtensor_ex}
\end{figure}

The next step is to compute the two-state QGT, which indicates all the states that are correlated after two parameter variations. To compute it, we use an approach similar to that in~\eqref{Q_Matrix}. Thus, we use~\eqref{M_def} and write
\begin{eqnarray}
	\label{M_matrix}
	\mathds{M}^{(m,n)} &:=&  \mfs{M^{(m,n)}_{ij\, MN}} = \sum_{l\neq m,n} \mfs{\sum_{L=1}^{d_{l}}  e^{(l,m)\,\ast}_{i\; LM} e^{(l,n)}_{j\; LN}}  = \sum_{l\neq m,n} \left( e^{(l,m)} \right)^{\dagger} \cdot  e^{(l, n)}.
\end{eqnarray}
Hence, we only need to determine which states $m$ are related to the initial state $n$ after two parameter variations. The simplest way to do it is to use the diagram in figure~\ref{Fig:Nbein_ex}; then, we overlap each black dot with another diagram centered on the point under study. Repeating this process for all the points in the original diagram results in figure~\ref{Fig:Mtensor_ex}. Here, the black markers, either circles or squares, represent all the states related to the blue state $(n_{1}, n_{d})$ by two parameter variations. Additionally, the circles are the points also related through one parameter variation (the $\mathcal{N}$-bein points); the red circle is only available through one parameter variation and not two.

From the diagram in figure~\ref{Fig:Mtensor_ex}, we notice several lines of states unreachable after parameter variations. The first is in $n_{d} \pm 1$, followed by another one at $n_{d}\pm 3$. If we consider a finite number of variations, these gaps will appear every two steps in $n_{d}$. It is a consequence from the fact that $e^{((n_{1}, n_{d} \pm 1), (n_{1}, n_{d}))}=0$, and it is a characteristic of the system under study highlighted by the use of $\mathcal{N}$-beins.

The sum in~\eqref{M_matrix} represents all the possible paths from one state to another. For example, to reach the point $(n_{1}+1, n_{d}+2)$, we can first go to $(n_{1}+1, n_{d})$ and then to $(n_{1}+1, n_{d}+2)$, whereas another path is $(n_{1}, n_{d}+2)$ then to $(n_{1}+1, n_{d}+2)$. In the former path, we first use the $\mathcal{N}$-bein $e^{((n_{1}+1, n_{d}),(n_{1}, n_{d}))}$ and then $e^{((n_{1}+1, n_{d}+2), (n_{1}+1, n_{d}))}$, while in the latter path the $\mathcal{N}$-bein order is from $e^{((n_{1}, n_{d}+2), (n_{1}, n_{d}))}$ to $e^{((n_{1}+1, n_{d}+2), (n_{1}, n_{d}+2))}$. Hence, to compute $\mathds{M}$ for every point in the grid of figure~\ref{Fig:Mtensor_ex}, we need to account for all the possible paths to arrive at it. Thus, we compute all the non-zero two-state tensors (one for each black marker in figure~\ref{Fig:Mtensor_ex}). Most of the two-state tensors are symmetric (the complete list is in appendix~\ref{Ap_Ex_extra}), so the torsion between these states is zero, and the order of variations is not important. Meanwhile, the only two-state tensors with an anti-symmetric part are:
\begin{eqnarray}
	\mathds{M}^{((n_{1}-1,n_{d}),(n_{1}, n_{d}))} & = & \frac{1}{8} \sqrt{\frac{3 n_{1}}{2 \hbar \omega_{1}^{7}}} \left[ \frac{n_{1}}{\omega_{1}^{2}} \begin{pmatrix}
		0 & \omega_{1}^{2} & 0 \\
		\omega_{1}^{2} & -2X & 0 \\
		0 & 0 & 0  
	\end{pmatrix} + \begin{pmatrix}
		0 & -1 & 0 \\
		1 & 0 & 0 \\
		0 & 0 & 0  
	\end{pmatrix} \right] \Kp \mathds{1}_{d_{n}}, \\
	\mathds{M}^{((n_{1}+1,n_{d}),(n_{1}, n_{d}))} & = & \frac{1}{8} \sqrt{\frac{3 (n_{1}+1)}{2 \hbar \omega_{1}^{7}}} \left[\frac{n_{1}+1}{\omega_{1}^{2}} \begin{pmatrix}
		0 & \omega_{1}^{2} & 0 \\
		\omega_{1}^{2} & -2X & 0 \\
		0 & 0 & 0  
	\end{pmatrix} - \begin{pmatrix}
		0 & -1 & 0 \\
		1 & 0 & 0 \\
		0 & 0 & 0  
	\end{pmatrix} \right] \Kp \mathds{1}_{d_{n}}. 
\end{eqnarray}
Therefore, the only non-zero torsion occurs when we go from the state to $(n_{1}, n_{d})$ to the states $(n_{1}\pm 1, n_{d})$, explicitly:
\begin{subequations}
	\begin{eqnarray}
		\label{T1_ex}
		\mathds{T}^{((n_{1}-1,n_{d}),(n_{1}, n_{d}))} & = & \frac{\mathrm{i}}{4} \sqrt{\frac{3 n_{1}}{2 \hbar \omega_{1}^{7}}}  \begin{pmatrix}
			0 & -1 & 0 \\
			1 & 0 & 0 \\
			0 & 0 & 0  
		\end{pmatrix} \Kp \mathds{1}_{d_{n}} \\
		\label{T2_ex}
		\mathds{T}^{((n_{1}+1,n_{d}),(n_{1}, n_{d}))} & = & - \frac{\mathrm{i}}{4} \sqrt{\frac{3 (n_{1}+1)}{2 \hbar \omega_{1}^{7}}}  \begin{pmatrix}
			0 & -1 & 0 \\
			1 & 0 & 0 \\
			0 & 0 & 0  
		\end{pmatrix}  \Kp \mathds{1}_{d_{n}}. 
	\end{eqnarray}
\end{subequations}

Recall that the torsion is also the covariant derivative of the $\mathcal{N}$-bein, see \eqref{T_De}. Consequently, if $e^{(m,n)}_{i\; MN}=0$ for a given state $\ket{m_{M}}$, then $T^{(m,n)}_{i\; MN}=0$. This means that only the black circles in figure~\ref{Fig:Mtensor_ex} can exhibit non-zero torsion, namely the states $m=(n_{1}\pm 1, n_{d})$ and $m=(n_{1}\pm 2, n_{d})$. Moreover, after two perturbations, the state $m=(n_{1}\pm 1, n_{d})$ is the only one that requires a reversal of direction along $n_{1}$ (increase and then decrease, or vice versa). At any other black marker, $n_{1}$ and $n_{d}$ increase or decrease or remain unchanged after two consecutive perturbations. 

To discuss the order sensitivity, consider the states with potential non-zero torsion, $m=(n_{1}\pm 1, n_{d})$ and $m=(n_{1}\pm 2, n_{d})$. A perturbation in $X$ induces one-unit jumps along $n_{1}$, while varying $k$ produces one- or two-unit jumps in $n_{1}$. To arrive at the states $m=(n_{1}\pm 2, n_{d})$, the admissible two-step sequences of perturbations are $XX$, $Xk$, $kX$, and $kk$. Hence, whether the first perturbation is $X$ or $k$, we are still able to reach $m=(n_{1}\pm 2, n_{d})$ when the second variation is $X$ or $k$. In this case, the order of perturbations is irrelevant. In contrast, to reach $m=(n_{1}\pm 1, n_{d})$, the admissible pairs are $Xk$, $kX$, and $kk$.  If the first perturbation is in $X$, then the second must be in $k$. Meanwhile, starting with $k$ does not constrain the second step. Here, the order of perturbations matters, and that is why this is the only transition with non-zero torsion.

\subsection{Invariants}

As we already mentioned, the importance of our formalism lies in the information given by the invariants that we derive from it. Before addressing the new invariants, let us discuss those prior to this work. Since $\mathds{Q}$ does not have an anti-symmetric part, it has real components. Given that $\Tr(\mathds{1}_{d_{n}})=d_{n}= n_{d} + 1$ and $\Tr[\mathds{B}(n_{d})]= (2/3) (n_{d}^{2} + 2 n_{d} +3)(n_{d}+1)$, we have
\begin{eqnarray}
	\label{TrQ}
	\Tr \mathds{Q}^{(n)} & = & \frac{3(2n_{1} + 1)(n_{d}+1)}{2 \hbar \omega_{1}^{7}}\begin{pmatrix}
		\omega_{1}^{4} & - \omega_{1}^{2} X & 0 \\
		- \omega_{1}^{2} X & X^{2} & 0 \\
		0 & 0 & 0  
	\end{pmatrix} 
	+ \frac{(n_{1}^{2} + n_{1} + 1)(n_{d} + 1)}{32 \omega_{1}^{4}}\begin{pmatrix}
		0 & 0 & 0 \\
		0 & 1 & 0 \\
		0 & 0 & 0  
	\end{pmatrix} \nonumber \\
	&& + \frac{(n_{d}^{2} + 2 n_{d} +3)(n_{d}+1)}{48 \omega_{2}^{4}}\begin{pmatrix}
		0 & 0 & 0 \\
		0 & 1 & 3 \\
		0 & 3 & 9  
	\end{pmatrix}.
\end{eqnarray}
Keep in mind that
\begin{equation}
	\Tr \mathds{Q}^{(n)} = \mfs{Q^{(n)}_{ij}} = \mfs{\sum_{N}^{d_{n}} Q^{(n)}_{ij \; NN} }.
\end{equation}
Thus, since the QGT is symmetric, its trace is equal to the metric
\begin{equation}
	\label{TrQg}
	\mfs{g^{(n)}_{ij}} = \Tr \mathds{Q}^{(n)}.
\end{equation}
From here, we derive more invariants, such as its determinant
\begin{equation}
	\det[g^{(n)}_{ij}] =\dfrac{9 (2n_{1}+1) (n_{1}^{2}+n_{1}+1) (n_{d}+1)^{3} (n_{d}^{2}+2n_{d}+3)}{1024 \hbar \omega_{1}^{7}\omega_{2}^{4}},
\end{equation}
and, for non-vanishing frequencies $\omega_{1}$ and $\omega_{2}$, its inverse metric
\begin{eqnarray}
	\mfs{g^{(n)\, ij}}& = & \frac{2 \hbar \omega_{1}^{3}}{3(2n_{1} + 1)(n_{d}+1)} \begin{pmatrix}
		1 & 0 & 0 \\
		0 & 0 & 0 \\
		0 & 0 & 0  
	\end{pmatrix} 
	+ \frac{16 \omega_{2}^{4}}{3 (n_{d}^{2} + 2 n_{d} +3)(n_{d}+1)}\begin{pmatrix}
		0 & 0 & 0 \\
		0 & 0 & 0 \\
		0 & 0 & 1  
	\end{pmatrix} \nonumber \\
	&& + \frac{32 }{9(n_{1}^{2} + n_{1} + 1)(n_{d} + 1)}\begin{pmatrix}
		9 X^{2} & 9 \omega_{1}^{2} X & -3 \omega_{1}^{2} X \\
		9 \omega_{1}^{2} X & 9 \omega_{1}^{4}  & - 3 \omega_{1}^{4}  \\
		-3 \omega_{1}^{2} X & -3\omega_{1}^{4} & \omega_{1}^{4} 
	\end{pmatrix}.
\end{eqnarray}
Notice how the determinant does not depend on the field strength $X$, only on frequencies and the quantum state $n=(n_{1}, n_{d})$. Moreover, the metric inverse allows us to compute the Riemann tensor and the scalar curvature for the parameter space; the latter is
\begin{equation}
	R(n)=-\frac{4}{(n_{1}^{2}+n_{1}+1)(n_{d}+1)}.
\end{equation}
Being a negative curvature, it is characteristic of a hyperbolic manifold, and as $n$ grows, the manifold tends to become flatter. Moreover, the curvature does not depend on any of the parameters, so in each $\mathcal{H}_{n}$ the curvature is constant and a parameter variation does not yield a change in the geometry of the manifold.

Among the different invariants we discussed, we are only going to compute three: the vector, the scalar, and the 3-form invariants \eqref{Inv_vec}, \eqref{Inv_NX}, and \eqref{inv_tau}, respectively. For the first case, we have a non-zero vector only when $n_{1}$ changes by at most one unit. Explicitly, the components of the only non-zero vectors are:
\begin{subequations}
	\begin{eqnarray}
		M^{((n_{1}-1,n_{d}),(n_{1},n_{d}))\, i}&=&  \frac{n_{1}+1}{n_{1}} \mathcal{G}^{((n_{1}-1,n_{d}),(n_{1},n_{d})) \, i} = \frac{n_{1}+1}{2 \mathrm{i}} T^{((n_{1}-1,n_{d}),(n_{1},n_{d})) \, i} \nonumber \\
		&=& -\frac{4 \mathrm{i} n_{1} (n_{1}+1)}{(2n_{1}+1)(n_{1}^{2}-n_{1}+1)(n_{d}+1)} V^{i},\\
		M^{((n_{1}+1,n_{d}),(n_{1},n_{d}))\, i}&=&  \frac{n_{1}}{n_{1}+1} \mathcal{G}^{((n_{1}+1,n_{d}),(n_{1},n_{d})) \, i} = - \frac{n_{1}}{2\mathrm{i}} T^{((n_{1}+1,n_{d}),(n_{1},n_{d})) \, i} \nonumber \\
		&=& \frac{4 \mathrm{i} n_{1} (n_{1}+1)}{(2n_{1}+1)(n_{1}^{2} + 3 n_{1} + 3)(n_{d}+1)} V^{i},
	\end{eqnarray}
\end{subequations}
where $V^{i}$ are the components of the vector
\begin{equation}
	\label{inv_vec_ex}
	V  := \dfrac{1}{3} \begin{pmatrix}
		3X \\
		3\omega_{1}^{2}\\ 
		-\omega_{1}^{2}
	\end{pmatrix}.
\end{equation}
Therefore, all of the vector invariants are proportional to each other, with the information encoded in the three-dimensional vector field $V$, which does not depend on the $k^{\prime}$ parameter---it is independent of the coupling. We interpret this vector as the direction on the parameter space to go from $\mathcal{H}_{(n_{1}, n_{d})}$ to $\mathcal{H}_{(n_{1} \pm 1, n_{d})}$, whether it depends on the parameter order or not. The vector $V$ does not depend on the coupling constant $k^{\prime}$ because the one-unit jumps are not related to variations on this parameter, see~\eqref{e1m} and \eqref{e1p}.

Continuing with the scalar invariants, we compute those in the form $\mathcal{N}_{\Xi}^{(m,n)}$ and associated with changes $m_{1}>n_{1}$ or $m_{d}>n_{d}$, since most of $\mathcal{N}_{\Xi \Theta}^{(m,n)}=0$ when $\Xi\neq \Theta$ and the cases $m_{1}<n_{1}$ or $m_{d}<n_{d}$ are equivalent due the property~\eqref{inv_sym}. The analytic expressions for the invariants are found in appendix \ref{Ap_Ex_extra}. All the scalar invariants depend only on the energy level $n$. Thus, for a fixed value $n_{d}=0$, we plot the results as a function of $n_{1}$. 
\begin{figure}[h]%
	\centering
	\subfloat[ ][]{\includegraphics[scale=0.9]{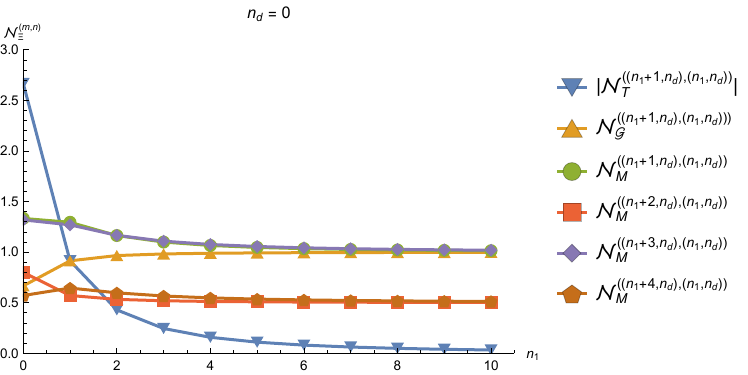}  \label{InvN_1}}
	\quad
	\subfloat[ ][]{\includegraphics[scale=0.9]{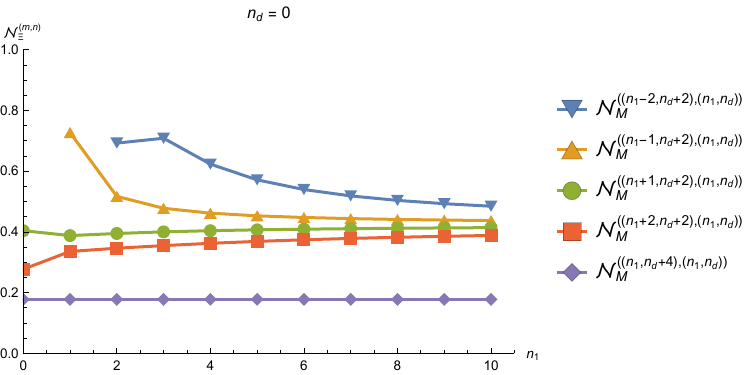}  \label{InvN_2}}
	\caption{Plots of some non-zero invariants $\mathcal{N}^{(m,n)}_{\Xi}$ for $M^{(m,n)}_{ij\, MN}$, $\mathcal{G}^{(m,n)}_{ij\, MN}$, and $T^{(m,n)}_{ij\, MN}$. (a) Graph of the invariants without degeneracy change, for $n_{d}=0$ and $\Xi=M, \, \mathcal{G}, \, T$. (b) Graph of the invariants with degeneracy change, for $n_{d}=0$ and $\Xi=M,\, \mathcal{G}, \, T$.}
\end{figure}
In figure~\ref{InvN_1}, we focus on the invariants associated with transitions that preserve the degeneracy $n_{d}$. For $n_{d}=0$, we recover the case reported in \cite{Romero2409}. In this figure, the invariant associated with the torsion ($\mathcal{N}^{(m,n)}_{T}$) has the largest magnitude moving from the ground state to the first excited level and is the only invariant that goes to zero as $n_{1}$ increases; the others approach a constant value. As noted, reaching the state $m=(n_{1}+1, n_{d})$ can occur via the perturbations $Xk$, $kX$, or $kk$. These sequences lead to the paths $(n_{1}, n_{d}) \rightarrow (n_{1}-1, n_{d}) \rightarrow (n_{1}+1, n_{d})$, $(n_{1}, n_{d}) \rightarrow (n_{1}+2, n_{d}) \rightarrow (n_{1}+1, n_{d})$, and either of the two, respectively. When $n_{1}=0$, only the second path is available. Consequently, the transition from $n=(0,n_{d})$ to $m=(1,n_{d})$ must begin varying the parameter $k$. This is where the order of variations is more significant, explaining why the torsion invariant reaches its maximum.

On the other hand,  in figure~\ref{InvN_2}, we plot the invariants whose state jumps induce a degeneracy change. In this case, we have zero torsion and $\mathcal{N}^{(m,n)}_{M}=\mathcal{N}^{(m,n)}_{\mathcal{G}}$. Moreover, we see in figure~\ref{InvN_2} that around the ground state, all the invariants have different values. However, as $n_{1}$ increases, most of them converge to a fixed value, see~\eqref{n1_limit2}. The only exception is $\mathcal{N}^{((n_{1},n_{d}+4),(n_{1},n_{d}))}_{M}$, whose value does not depend on $n_{1}$, see~\eqref{NMndp4}.

Finally, we use~\eqref{inv_tau}, without section~\ref{Sec_df} conventions, to compute the 3-form invariant $\tau^{(n)}$, namely
\begin{eqnarray}
	\boldsymbol{\tau}^{(n)} &=& \sum_{N=1}^{d_{n}} \sum_{m\neq n} \sum_{M=1}^{d_{m}} T^{(m,n)}_{i_{1} i_{2} \; MN} e^{(n,m)}_{i_{3} \; NM} \; d\lambda^{i_{1}} \wedge d\lambda^{i_{2}} \wedge d\lambda^{i_{3}} \nonumber \\
	&=& \sum_{N=1}^{d_{n}} \sum_{M=1}^{d_{n}} \Big( T^{((n_{1}-1,n_{d}), (n_{1}, n_{d}))}_{i_{1} i_{2} \; MN} e^{((n_{1},n_{d}), (n_{1}-1, n_{d}))}_{i_{3} \; NM} \nonumber \\
	&& + T^{((n_{1}+1,n_{d}), (n_{1}, n_{d}))}_{i_{1} i_{2} \; MN} e^{((n_{1},n_{d}), (n_{1}+1, n_{d}))}_{i_{3} \; NM} \Big) d\lambda^{i_{1}} \wedge d\lambda^{i_{2}} \wedge d\lambda^{i_{3}} 
	= 0.
\end{eqnarray}
Since the dimension of our parameter space is three, the invariant $\boldsymbol{\tau}^{(n)}$ is proportional to the volume form $dX \wedge dk \wedge dk^{\prime}$. The vanishing proportionality constant implies that either the contributions for all the terms in the sum over $m\neq n$ cancel out, or that each term is identically zero. In our case, it is the latter instance, because the $\mathcal{N}$-bein and the torsion lie in the same plane, the $Xk$-plane. Since $\boldsymbol{\tau}^{(n)}$ is a 3-form, it is associated with third-order variations. The fact that $\boldsymbol{\tau}^{(n)}=0$ is due to the simplicity of our example. We did not expect that third-order corrections were necessary for this system.

\section{Conclusions}
\label{Sec_concl}

In this work, we extend the formalism introduced in~\cite{Romero2409} by generalizing the $\mathcal{N}$-bein for quantum systems with degenerate spectra within Cartan's geometry. In the process, we explore the consequences of the additional symmetries of degenerate Hilbert spaces and study how they shape the geometry of the parameter space of quantum mechanics. 

Our central innovation is the generalized $\mathcal{N}$-bein, $e^{(m,n)}_{i\; MN}$. While the standard WZ connection operates within a single degenerate subspace, our generalized 1-form bridges states across different subspaces ($\mathcal{H}_{n}$) that may possess entirely different degeneracies. Transforming as a rectangular matrix under gauge transformations, it strictly encodes the possibility of transitioning from a state $\ket{n_{N}}$ to another $\ket{m_{M}}$ following an infinitesimal perturbation of the parameter $\lambda^{i}$. A vanishing $\mathcal{N}$-bein ($e^{(m,n)}_{i\; MN}=0$) explicitly dictates that the target state is unreachable via that specific parameter variation. 

Using the generalized $\mathcal{N}$-bein, we defined a non-Abelian two-state QGT $M^{(m,n)}_{ij\, MN}$, which identifies possible transitions from $\ket{n_{N}}$ to $\ket{m_{M}}$ after two successive parameter variations. Its symmetric part $\mathcal{G}^{(m,n)}_{ij\, MN}$ determines when the order of parameter variations is irrelevant. In contrast, the anti-symmetric part $T^{(m,n)}_{ij\, MN}$ indicates when the order of variations matters. We call the latter torsion because it is equal to the covariant derivative of the $\mathcal{N}$-bein. Thus, the correspondence between the Wilczek–Zee connection and its curvature with the anti-symmetric part of the standard QGT finds its analogue in the relation between the $\mathcal{N}$-bein and torsion with the anti-symmetric part of the two-state QGT.

For degenerate spectra, the Hilbert space decomposes as subspaces $\mathcal{H}_{n}$, each one associated with an energy $E_{n}$. Unlike the WZ connection, the $\mathcal{N}$-bein contains information about two states that lie in different subspaces (potentially with different degeneracies). Nonetheless, we derive the transformation laws of all the two-state tensors under a change of basis in each $\mathcal{H}_{n}$. Although the $\mathcal{N}$-bein and the two-state tensors are not invariant under the transformation, we use them to construct gauge invariants; most notably, the square of the $\mathcal{N}$-bein~\eqref{qij_def}, a vector~\eqref{Inv_vec}, a scalar~\eqref{Inv_scl}, and a 3-form~\eqref{inv_tau}. The invariants serve as key ingredients to build quantum observables. In particular, the 3-form is integrable in a three-dimensional parameter manifold and generates a new topological invariant of the Nieh-Yan type \cite{Nieh}, see also \cite{Chandia}. Whether this invariant may be linked to a new type of conductivity is left for future research. 

Presenting our formalism using differential forms exhibits the geometric nature behind the objects under consideration, justifying their definitions. This notation simplifies the expressions and finds parallels with different non-Abelian gauge theories. In particular, just as the curvature and the WZ connection satisfy the Bianchi identity \eqref{B1}, the $\mathcal{N}$-bein and the torsion satisfy \eqref{B2}.

In the example of section~\ref{Sec_ejem}, we use our formalism to expose invariants with topological information, Figs.~\ref{InvN_1} and~\ref{InvN_2}. These invariants were constructed using two-state tensors and depend solely on the quantum state under consideration, thereby revealing parameter-independent features. The invariants exhibit significant sensitivity to states near the ground state, which may be relevant for their measurement in such quantum systems. Furthermore, these characteristics are relevant to holonomic quantum computation because they could be invariants that are protected against variations in the system parameters.

In summary, the formalism developed here provides a unified geometric toolkit ($\mathcal{N}$-bein, two-state geometric tensor, torsion, and the associated invariants) to study degenerate quantum systems. We hope it will facilitate both conceptual understanding and practical applications. Explicitly, in the study of quantum materials \cite{Mera}, quantum metrology \cite{Fadel}, quantum optics \cite{Ahn}, flat-band superconductivity \cite{Reko}, among others.

\section*{Acknowledgements}
J.R. acknowledges the financial support from Secihti under the ``Estancias Posdoctorales por M\'{e}xico 2022 (3)'' program. C. A. V. gratefully acknowledges Secihti for his PhD scholarship (662129).

\section*{Data availability statement}
All data that support the findings of this study are included within the article (and any supplementary files).

\section*{Funding}
This work was partially supported by DGAPA-PAPIIT Grant No. IN114225.

\bibliographystyle{unsrt}
\bibliography{References}

\appendix
\section{Properties of the geometric objects}
\label{Ap_prop}

Given the numerous two-state tensors and invariants we have defined, we dedicate this part to elaborating on the different properties they possess. Apart from the tensor symmetries in the parameter indices, the two most relevant qualities are their change under complex conjugation and under the interchange of quantum states. These properties are intertwined, and they are worth discussing.

\subsection{Tensor properties}

We begin with the $\mathcal{N}$-bein, from its definition  
\begin{equation}
	e^{(m,n)}_{i\; MN} := \mathrm{i} \bk{m_{M}}{\partial_{i} n_{N}},
\end{equation}	
and the identity from the normalization condition~\eqref{norm}, 
\begin{equation}
    \label{e_norm_id}
	\bk{\partial_{i} m_{M}}{n_{N}} + \bk{m_{M}}{\partial_{i} n_{N}} = 0,
\end{equation}
it is straightforward to prove
\begin{equation}
	\label{conj}
	e^{(m,n)\; \ast}_{i\; MN}  = e^{(n,m)}_{i\; NM}. 
\end{equation}
The complex conjugation of the $\mathcal{N}$-bein acts as an exchange of quantum states. Since a non-zero $\mathcal{N}$-bein indicates a possible transition from $\ket{n_{N}}$ to $\ket{m_{M}}$, its conjugate corresponds to the reverse transition, from $\ket{m_{M}}$ back to $\ket{n_{N}}$. Consequently, whenever a parameter variation allows the system to reach a given state, the same variation ensures the possibility of returning to the initial state. In this way, we have a symmetry between states through parameter variations.

From the last identity and the QGT definition~\eqref{qgt_def}, we derive the property
\begin{equation}
	Q^{(n)\; \ast}_{ij\; N_{1} N_{2}} = Q^{(n)}_{ji\; N_{2} N_{1}}.
\end{equation}
Thus, since 
\begin{equation}
	Q^{(n)}_{ij\; N_{1} N_{2}} = g^{(n)}_{ij\; N_{1} N_{2}} + \dfrac{1}{2 \mathrm{i}} F^{(n)}_{ij\; N_{1} N_{2}},
\end{equation}
we have
\begin{subequations}
	\begin{eqnarray}
		g^{(n)\; \ast}_{ij\; N_{1} N_{2}} &=& g^{(n)}_{ij\; N_{2} N_{1}}, \\
		F^{(n)\; \ast}_{ij\; N_{1} N_{2}} &=& F^{(n)}_{ij\; N_{2} N_{1}}.
	\end{eqnarray}
\end{subequations}
Hence, the elements of these tensors, $g^{(n)}_{ij\; N_{1} N_{2}}$ and $F^{(n)}_{ij\; N_{2} N_{1}}$, are real for  $N_{1}=N_{2}$. Moreover, 	the invariants $g^{(n)}_{ij}$ and $F^{(n)}_{ij}$---defined in \eqref{gij_def} and \eqref{Fij_def}, respectively--- are real gauge invariants.

On the other hand, we have a similar behavior for the two-state QGT
\begin{equation}
	\label{M_conj}
	M^{(m,n)\; \ast}_{ij\; MN} = M^{(n, m) }_{ji\; NM}.
\end{equation}
And from
\begin{equation}
	\label{M_GT}
	M^{(m,n)}_{ij\; MN} = \mathcal{G}^{(m,n)}_{ij\; MN} + \dfrac{1}{2 \mathrm{i}} T^{(m,n)}_{ij\; MN},
\end{equation}
we obtain
\begin{subequations}
	\begin{eqnarray}
		\label{G_conj}
		\mathcal{G}^{(m,n)\; \ast}_{ij\; MN} &=& \mathcal{G}^{(n,m)}_{ij\; NM}, \\
		\label{T_conj}
		T^{(m,n)\; \ast}_{ij\; MN} &=& T^{(n,m)}_{ij\; NM}.
	\end{eqnarray}
\end{subequations}
Therefore, for $\mathcal{G}^{(m,n)}_{ij\; MN}$ and $T^{(m,n)}_{ij\; MN}$, the conjugation operation does not alter the parameter indices and changes the state order, similar to the $\mathcal{N}$-bein. If after two parameter variations we can reach the state $\ket{m_{M}}$ from $\ket{n_{N}}$, then it is also possible to go from $\ket{m_{M}}$ to $\ket{n_{N}}$. This property is inherited by the $\mathcal{N}$-bein.

\subsection{Invariants properties}
\label{Ap_Sec_inv}

The gauge invariants represent a key aspect of the formalism; with them, we construct observables derived from the two-state tensors. Thus, we decided to go into further detail to explore their properties. The first invariants we discussed were the trace of the QGT $Q^{(n)}_{ij}$ and its real and imaginary parts, given by \eqref{Qij_def}, \eqref{gij_def}, and \eqref{Fij_def}, respectively. These invariants were introduced prior to our work and are defined for each subspace $\mathcal{H}_{n}$. The latter two are real and are the trace of the quantum geometric metric and the WZ curvature.  

Regarding the two-state invariants, the first ones we showed are the squares of the $\mathcal{N}$-bein, which are also the summands that compose the $Q^{(n)}_{ij}$, see \eqref{Qij_qij}, namely
\begin{equation}
	q^{(n,m)}_{ij} := \sum_{N=1}^{d_{n}} \sum_{M=1}^{d_{m}} e^{(n,m)}_{i\; NM} e^{(m,n)}_{j\; MN}.
\end{equation} 
Given the conjugation property of the $\mathcal{N}$-bein, the summands satisfy
\begin{equation}
	q^{(n,m)\; \ast}_{ij}  = q^{(m,n)}_{ij} = q^{(n,m)}_{ji}.
\end{equation}
Thus, a complex conjugation in $q^{(n,m)}_{ij}$ is equivalent to the exchange of quantum states or the exchange of parameter indices, but not both simultaneously. Hence, if $q^{(n,m)}_{ij}$ is symmetric, then $q^{(n,m)}_{ij}$ is real and vice versa, just like the QGT. 

On the other hand, the properties of the two-state-tensor invariants---either with one \eqref{Inv_vec}, three \eqref{inv_R}, or four \eqref{inv_S} parameter indices---depend on the tensors used to construct them. Among different combinations of two-state invariants, the more interesting properties arise when $\Xi^{(m,n)}_{ij \; MN}$ and $\Theta^{(m,n)}_{ij \; MN}$ equal $\mathcal{G}^{(m,n)}_{ij \; MN}$ or $T^{(m,n)}_{ij \; MN}$. In this case, we have
\begin{subequations}
	\begin{eqnarray}
		\left( R_{\Xi} \right)^{(m,n)\, \ast}_{ijk} &=& \left( R_{\Xi} \right)^{(n,m)}_{ijk}, \\
		\left( S_{\Xi \Theta} \right)^{(m,n)\, \ast}_{ijkl} &=& \left( S_{\Xi \Theta} \right)^{(n,m)}_{ijkl}.
	\end{eqnarray}
\end{subequations}
Here, the conjugation exchanges the states, similar to some of the two-state tensors. However, in this case we are certain that the property holds after a gauge transformation. Moreover, from \eqref{G_conj} and \eqref{T_conj} we derive the properties:
\begin{subequations}
	\begin{eqnarray}
		\mathrm{Re} \left\lbrace \Xi^{(m,n)}_{ij\; MN} \right\rbrace & = & \mathrm{Re} \left\lbrace \Xi^{(n,m)}_{ij\; NM} \right\rbrace,\\
		\mathrm{Im} \left\lbrace \Xi^{(m,n)}_{ij\; MN} \right\rbrace & = & -\mathrm{Im} \left\lbrace \Xi^{(n,m)}_{ij\; NM} \right\rbrace.
	\end{eqnarray}
\end{subequations}
Therefore, we rewrite the real and imaginary parts of $\left( S_{\Xi \Theta} \right)^{(m,n)}_{ijkl}$ as
\begin{subequations}
	\begin{eqnarray}
		\left( N_{\Xi \Theta} \right)^{(m,n)}_{ijkl} &:=& \sum_{N=1}^{d_{n}} \sum_{M=1}^{d_{m}} \left( \mathrm{Re} \left\lbrace \Xi^{(m,n)}_{ij\; MN} \right\rbrace \mathrm{Re} \left\lbrace \Theta^{(n,m)}_{kl\; NM} \right\rbrace - \mathrm{Im} \left\lbrace \Xi^{(m,n)}_{ij\; MN} \right\rbrace \mathrm{Im} \left\lbrace \Theta^{(n,m)}_{kl\; NM} \right\rbrace \right)  \\
		& = & \sum_{N=1}^{d_{n}} \sum_{M=1}^{d_{m}} \left( \mathrm{Re} \left\lbrace \Xi^{(m,n)}_{ij\; MN} \right\rbrace \mathrm{Re} \left\lbrace \Theta^{(m,n)}_{kl\; MN} \right\rbrace + \mathrm{Im} \left\lbrace \Xi^{(m,n)}_{ij\; MN} \right\rbrace \mathrm{Im} \left\lbrace \Theta^{(m,n)}_{kl\; MN} \right\rbrace \right) , \\
		\left( A_{\Xi \Theta} \right)^{(m,n)}_{ijkl} &:=&  \sum_{N=1}^{d_{n}} \sum_{M=1}^{d_{m}} \left( \mathrm{Re} \left\lbrace \Xi^{(m,n)}_{ij\; MN} \right\rbrace \mathrm{Im} \left\lbrace \Theta^{(n,m)}_{kl\; NM} \right\rbrace + \mathrm{Im} \left\lbrace \Xi^{(m,n)}_{ij\; MN} \right\rbrace \mathrm{Re} \left\lbrace \Theta^{(n,m)}_{kl\; NM} \right\rbrace \right) \\
		& = & - \sum_{N=1}^{d_{n}} \sum_{M=1}^{d_{m}} \left( \mathrm{Re} \left\lbrace \Xi^{(m,n)}_{ij\; MN} \right\rbrace \mathrm{Im} \left\lbrace \Theta^{(m,n)}_{kl\; MN} \right\rbrace - \mathrm{Im} \left\lbrace \Xi^{(m,n)}_{ij\; MN} \right\rbrace \mathrm{Re} \left\lbrace \Theta^{(m,n)}_{kl\; MN} \right\rbrace \right). \qquad
	\end{eqnarray}
\end{subequations}
The invariants $\left( N_{\Xi \Theta} \right)^{(m,n)}_{ijkl}$ and $\left( A_{\Xi \Theta} \right)^{(m,n)}_{ijkl}$ are real; hence, they are viable to be observables. Also, under a state exchange, they satisfy;
\begin{subequations}
	\begin{eqnarray}
		\left( N_{\Xi \Theta} \right)^{(m,n)}_{ijkl} &=& \left( N_{\Xi \Theta} \right)^{(n,m)}_{ijkl},\\
		\left( A_{\Xi \Theta} \right)^{(m,n)}_{ijkl} &=& -\left( A_{\Xi \Theta} \right)^{(n,m)}_{ijkl}.
	\end{eqnarray}
\end{subequations}
Recall that we are only consider the tensors $\mathcal{G}^{(m,n)}_{ij \; MN}$ or $T^{(m,n)}_{ij \; MN}$, for $M^{(m,n)}_{ij\; MN}$ the properties change a little due to the indices swap in \eqref{M_conj}. Nevertheless, in the limit where there is no degeneracy ($d_{n}=1$, for all $n$), the invariants presented here are equivalent to those for the non-degenerate case \cite{Romero2409}.

Finally, we discuss the scalar invariant. There exist two different ways to contract the indices with the metric inverse, one being in a cyclic contraction and the other being similar to the Ricci scalar curvature. Furthermore, there are also two distinct ways to construct an invariant. The first is with the transposition of the indices, and the second is with the conjugation. When the tensor is $\mathcal{G}^{(m,n)}_{i j \; MN}$ or $T^{(m,n)}_{i j \; MN}$, both of these operations are the same, see~\eqref{G_conj} and \eqref{T_conj}. However, they are different for the two-state tensor $M^{(m,n)}_{i j \; MN}$, see~\eqref{M_conj}. Taking into account the two possible contractions and the two operations, there are four different ways to construct the scalar invariant; they are:	
\begin{itemize}
	\item[\emph{a)}] Cyclic contraction + Transposition
	
	Scalar invariant definition
	\begin{equation}
		\label{inv_a}
		\mathcal{N}_{\Xi \Theta}^{(m,n)} := 2 \sum_{N=1}^{d_{n}} \sum_{M=1}^{d_{m}} g^{(m)\, ij} \Xi^{(m,n)}_{jk\; MN}  g^{(n)\, kl} \Theta^{(n,m)}_{li\; NM},
	\end{equation}
	Identity between the tensors
	\begin{equation}
		\mathcal{N}_{M}^{(m,n)} =  \mathcal{N}_{\mathcal{G}}^{(m,n)} - \frac{1}{4} \mathcal{N}_{T}^{(m,n)} + \mathrm{Im} \left\lbrace \mathcal{N}_{\mathcal{G}T}^{(m,n)} \right\rbrace.
	\end{equation}

	\item[\emph{b)}] Cyclic contraction + Conjugate
	
	Scalar invariant definition
	\begin{equation}
		\label{inv_b}
		\mathcal{N}_{\Xi \Theta}^{(m,n)} := 2 \sum_{N=1}^{d_{n}} \sum_{M=1}^{d_{m}} g^{(m)\, ij} \Xi^{(m,n)}_{jk\; MN}  g^{(n)\, kl} \Theta^{(m,n)\; \ast}_{li\; MN},
	\end{equation}
	Identity between the tensors
	\begin{equation}
		\mathcal{N}_{M}^{(m,n)} =  \mathcal{N}_{\mathcal{G}}^{(m,n)} + \frac{1}{4} \mathcal{N}_{T}^{(m,n)} + \mathrm{i} \mathrm{Re} \left\lbrace \mathcal{N}_{\mathcal{G}T}^{(m,n)} \right\rbrace.
	\end{equation}
	
	\item[\emph{c)}] Ricci-like contraction + Transposition
	
	Scalar invariant definition
	\begin{equation}
		\label{inv_c}
		\mathcal{N}_{\Xi \Theta}^{(m,n)} := 2 \sum_{N=1}^{d_{n}} \sum_{M=1}^{d_{m}} g^{(m)\, ij} g^{(n)\, kl} \Xi^{(m,n)}_{ik\; MN} \Theta^{(n,m)}_{jl \; NM},
	\end{equation}
	Identity between the tensors
	\begin{equation}
		\mathcal{N}_{M}^{(m,n)} =  \mathcal{N}_{\mathcal{G}}^{(m,n)} - \frac{1}{4} \mathcal{N}_{T}^{(m,n)} -\mathrm{i} \mathrm{Re} \left\lbrace \mathcal{N}_{\mathcal{G}T}^{(m,n)} \right\rbrace.
	\end{equation}

	\item[\emph{d)}] Ricci-like contraction + Conjugation
	
	Scalar invariant definition
	\begin{equation}
		\label{inv_d}
		\mathcal{N}_{\Xi \Theta}^{(m,n)} := 2 \sum_{N=1}^{d_{n}} \sum_{M=1}^{d_{m}} g^{(m)\, ij} g^{(n)\, kl} \Xi^{(m,n)}_{ik\; MN} \Theta^{(m,n)\; \ast}_{jl \; MN},
	\end{equation}
	Identity between the tensors
	\begin{equation}
		\mathcal{N}_{M}^{(m,n)} =  \mathcal{N}_{\mathcal{G}}^{(m,n)} + \frac{1}{4} \mathcal{N}_{T}^{(m,n)} - \mathrm{Im} \left\lbrace \mathcal{N}_{\mathcal{G}T}^{(m,n)} \right\rbrace.
	\end{equation}
	
\end{itemize}

All of the definitions for the scalar invariants have similar properties, and some of them are equivalent for certain tensors given their symmetry and conjugation properties. However, for this work we decided to use the first case (Cyclic contraction + Transposition), which differs from that given in~\cite{Romero2409}.  One of the advantages of this definition is that the invariant
\begin{equation}
	\mathcal{N}_{\Xi}^{(m,n)} = 2 \sum_{N=1}^{d_{n}} \sum_{M=1}^{d_{m}} g^{(m)\, ij} \Xi^{(m,n)}_{jk\; MN}  g^{(n)\, kl} \Xi^{(n,m)}_{li\; NM},
\end{equation}
is always a real function of the parameters for the three possible values of $\Xi$. Also, it fulfills the symmetric property between states
\begin{equation}
	\mathcal{N}_{\Xi}^{(m,n)} = \mathcal{N}_{\Xi}^{(n,m)}.
\end{equation}
Similarly, with the definition~\eqref{inv_a} we find the properties
\begin{subequations}
	\begin{eqnarray}
		\label{N_MG}
		\mathcal{N}_{M \mathcal{G}}^{(m,n)} & = & \mathcal{N}_{\mathcal{G} M}^{(n,m)}  = \mathcal{N}_{M \mathcal{G}}^{(n,m) \, \ast},\\
		\label{N_MT}
		\mathcal{N}_{MT}^{(m,n)} & = & \mathcal{N}_{TM}^{(n,m)}  = - \mathcal{N}_{MT}^{(n,m) \, \ast},\\
		\label{N_GT}
		\mathcal{N}_{\mathcal{G} T}^{(m,n)} & = & \mathcal{N}_{T \mathcal{G}}^{(n,m)}  = - \mathcal{N}_{\mathcal{G} T}^{(n,m) \, \ast}.
	\end{eqnarray}
\end{subequations}

In general, the scalar invariant is a function of the parameters that characterize a quantum system. It is an observable between two different Hilbert subspaces $\mathcal{H}_{m}$ and $\mathcal{H}_{n}$. In the case where it does not depend on the parameters, it becomes a property between the states of the system under study.\\

\section{Additional example content}
\label{Ap_Ex_extra}

In section~\ref{Sec_ejem}, we applied our formalism to the case of a three-dimensional harmonic oscillator subjected to an external electric field. While we outlined the general procedure for computing the relevant tensors, there remain several technical subtleties that merit further discussion.

\subsection{$\mathcal{N}$-bein visualization}
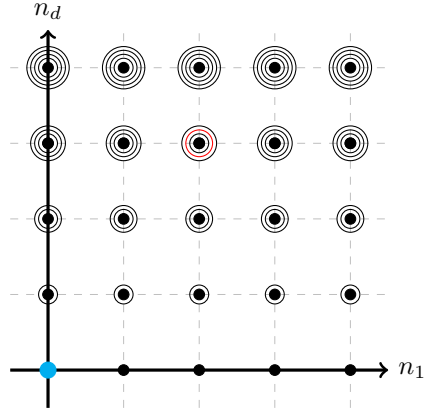
\begin{figure}[h!]
	\centering
	\begin{tikzpicture}
		
		\draw[very thin, color=gray!50, dashed] (-0.5,-0.5) grid (4.5,4.5);
		
		\draw[->, very thick] (-0.5,0) -- (4.5,0) node[right] {$n_{1}$};
		\draw[->, very thick] (0,-0.5) -- (0,4.5) node[above] {$n_{d}$};
		
		\filldraw[cyan] (0,0) circle (3pt);
		\filldraw[black] (1,0) circle (2pt);
		\filldraw[black] (2,0) circle (2pt);
		\filldraw[black] (3,0) circle (2pt);
		\filldraw[black] (4,0) circle (2pt);
		
		\filldraw[black] (0,1) circle (2pt);
		\filldraw[black] (1,1) circle (2pt);
		\filldraw[black] (2,1) circle (2pt);
		\filldraw[black] (3,1) circle (2pt);
		\filldraw[black] (4,1) circle (2pt);
		
		\filldraw[black] (0,2) circle (2pt);
		\filldraw[black] (1,2) circle (2pt);
		\filldraw[black] (2,2) circle (2pt);
		\filldraw[black] (3,2) circle (2pt);
		\filldraw[black] (4,2) circle (2pt);
		
		\filldraw[black] (0,3) circle (2pt);
		\filldraw[black] (1,3) circle (2pt);
		\filldraw[black] (2,3) circle (2pt);
		\filldraw[black] (3,3) circle (2pt);
		\filldraw[black] (4,3) circle (2pt);
		
		\filldraw[black] (0,3) circle (2pt);
		\filldraw[black] (1,3) circle (2pt);
		\filldraw[black] (2,3) circle (2pt);
		\filldraw[black] (3,3) circle (2pt);
		\filldraw[black] (4,3) circle (2pt);
		
		\filldraw[black] (0,4) circle (2pt);
		\filldraw[black] (1,4) circle (2pt);
		\filldraw[black] (2,4) circle (2pt);
		\filldraw[black] (3,4) circle (2pt);
		\filldraw[black] (4,4) circle (2pt);
		
		\draw (0,1) circle (0.125);
		\draw (1,1) circle (0.125);
		\draw (2,1) circle (0.125);
		\draw (3,1) circle (0.125);
		\draw (4,1) circle (0.125);

		\draw (0,2) circle (0.125);
		\draw (1,2) circle (0.125);
		\draw (2,2) circle (0.125);
		\draw (3,2) circle (0.125);
		\draw (4,2) circle (0.125);
		
		\draw (0,2) circle (0.177);
		\draw (1,2) circle (0.177);
		\draw (2,2) circle (0.177);
		\draw (3,2) circle (0.177);
		\draw (4,2) circle (0.177);

		\draw (0,3) circle (0.125);
		\draw (1,3) circle (0.125);
		\draw (2,3) circle (0.125);
		\draw (3,3) circle (0.125);
		\draw (4,3) circle (0.125);
		
		\draw (0,3) circle (0.177);
		\draw (1,3) circle (0.177);
		\draw[red] (2,3) circle (0.177);
		\draw (3,3) circle (0.177);
		\draw (4,3) circle (0.177);
		
		\draw (0,3) circle (0.23);
		\draw (1,3) circle (0.23);
		\draw (2,3) circle (0.23);
		\draw (3,3) circle (0.23);
		\draw (4,3) circle (0.23);

		\draw (0,4) circle (0.125);
		\draw (1,4) circle (0.125);
		\draw (2,4) circle (0.125);
		\draw (3,4) circle (0.125);
		\draw (4,4) circle (0.125);
		
		\draw (0,4) circle (0.177);
		\draw (1,4) circle (0.177);
		\draw (2,4) circle (0.177);
		\draw (3,4) circle (0.177);
		\draw (4,4) circle (0.177);
		
		\draw (0,4) circle (0.23);
		\draw (1,4) circle (0.23);
		\draw (2,4) circle (0.23);
		\draw (3,4) circle (0.23);
		\draw (4,4) circle (0.23);
		
		\draw (0,4) circle (0.28);
		\draw (1,4) circle (0.28);
		\draw (2,4) circle (0.28);
		\draw (3,4) circle (0.28);
		\draw (4,4) circle (0.28);
		
	\end{tikzpicture}
	\caption{Diagram showing different states $\ket{n_{N}}$. Each dot represents an energy level, with the blue dot being the ground state. The rings surrounding each black dot account for a different value of $N$. As $N$ increases, we move to the outer rings, with $N=1$ being the black dot. The state $\ket{n_{N}}$ with $n=(2,3)$ and $N=3$ is highlighted in red.}
	\label{Fig:states_2}
\end{figure}

In the example, we found that the Hilbert space is composed of subspaces $\mathcal{H}_{n}$ with energy $E_{n}$; see \eqref{En}. Here, the quantum numbers $n_{1}$ and $n_{d}$ characterize each energy level, $n=(n_{1}, n_{d})$, where the degeneracy on each subspace is $d_{n}=n_{d}+1$. Although figure~\ref{Fig:states} represents each state with energy $E_{n}$ as a point, such a diagram does not account for the degeneracy. To help us visualize it, we use the diagram in figure~\ref{Fig:states_2}. Again, each point corresponds to a given energy $E_{n}$. Additionally, the rings surrounding each dot represent a different value of $N$. When $N=1$, we are at the center, the black dot. As we increase the value of $N$, we move to the outer rings, and the number of rings is limited by the value of $d_{n}$. In figure~\ref{Fig:states_2}, we highlighted the state $\ket{n_{N}}=\ket{(2,3)_3}$ with a red ring. Furthermore, notice that all the levels along the lines $n_{d}=constant$ have the same number of rings; they are all the states with the same degeneracy. 		

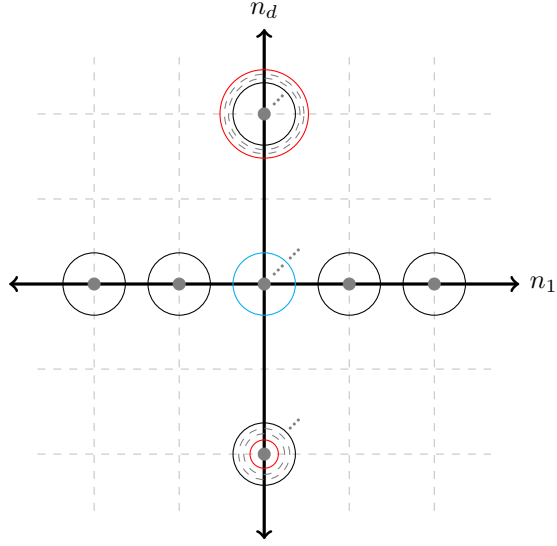
\begin{figure}[h!]
	\centering
	\begin{tikzpicture}[scale=0.75]
		
		\draw[very thin, color=gray!50, dashed,step=1.5] (-4,-4) grid (4,4);
		
		\draw[<->, very thick] (-4.5,0) -- (4.5,0) node[right] {$n_{1}$};
		\draw[<->, very thick] (0,-4.5) -- (0,4.5) node[above] {$n_{d}$};
		
		\filldraw[gray] (0,0) circle (3pt);
		\filldraw[gray] (1.5,0)  circle (3pt);
		\filldraw[gray] (-1.5,0) circle (3pt);
		\filldraw[gray] (3,0) circle (3pt);
		\filldraw[gray] (-3,0)  circle (3pt);
		\filldraw[gray] (0,3) circle (3pt);
		\filldraw[gray] (0,-3) circle (3pt);
		
		\filldraw [gray] (45:0.25) circle (0.5pt);
		\filldraw [gray] (45:0.35) circle (0.5pt);
		\filldraw [gray] (45:0.45) circle (0.5pt);
		
		\filldraw [gray] (45:0.65) circle (0.5pt);
		\filldraw [gray] (45:0.75) circle (0.5pt);
		\filldraw [gray] (45:0.85) circle (0.5pt);
		
		
		\draw[cyan] (0,0) circle (0.55);
		\draw[black] (-3,0) circle (0.55);
		\draw[black] (-1.5,0) circle (0.55);
		\draw[black] (1.5,0) circle (0.55);
		\draw[black] (3,0) circle (0.55);
		
		\draw[black] (0,3) circle (0.55);
		\draw[gray, dashed] (0,3) circle (0.63);
		\draw[gray, dashed] (0,3) circle (0.7);
		\draw[red] (0,3) circle (0.78);
		
		\filldraw [gray] (0,3) ++ (45:0.25) circle (0.5pt);
		\filldraw [gray] (0,3) ++ (45:0.35) circle (0.5pt);
		\filldraw [gray] (0,3) ++ (45:0.45) circle (0.5pt);

		\draw[red] (0,-3) circle (0.25);
		\draw[gray, dashed] (0,-3) circle (0.36);
		\draw[gray, dashed] (0,-3) circle (0.45);
		\draw[black] (0,-3) circle (0.55);
		
		\filldraw [gray] (0,-3) ++ (45:0.65) circle (0.5pt);
		\filldraw [gray] (0,-3) ++ (45:0.75) circle (0.5pt);
		\filldraw [gray] (0,-3) ++ (45:0.85) circle (0.5pt);
		
	\end{tikzpicture}
	\caption{Diagram showing all the states (with their degeneracy) related through a parameter variation with the initial state $n=(n_{1}, n_{d})$, blue ring. The solid black rings represent all the states with the same value $N$, while the solid red lines represent a jump of two units, either $N+2$ (top ring) or $N-2$ (bottom ring).}
	\label{Fig:Nbein_ex_2}
\end{figure}

On the other hand, in figure~\ref{Fig:Nbein_ex} we showed all the states related to the initial state $n=(n_{1}, n_{d})$ through a parameter variation, i.e., we are plotting all non-zero $\mathcal{N}$-beins. However, from \eqref{e3m} and \eqref{e3p}, we recognize two possible jumps that change the $N$ direction, either to $N-2$ or to $N+2$. Therefore, a variation on the parameters (particularly $k$ or $k^{\prime}$) could move the state to one with higher or lower degeneracy and, at the same time, change the direction in the new subspace $\mathcal{H}_{(n_{1}, n_{d}\pm 2)}$. In figure~\ref{Fig:Nbein_ex_2}, the initial state $\ket{n_{N}}$ is the blue ring at the center. All the possible state changes that do not change degeneracy are along the $n_{1}$ axis. They do not change the value of $N$ and correspond to the solid black rings centered at the valid $n_{1}$ values (gray dots). When we change to a state with different degeneracy $(n_{1}, n_{d} \pm 2$), we have two options: either we remain in the same direction $N$ [black rings centered at ($n_{1}, n_{d}\pm 2$)] or we move to $N\pm2$ (solid red rings).  

\subsection{Two-state tensor}

Consider the state $\ket{(n_{1}, n_{d})_{N}}$. After two consecutive parameter variations, we end up in one of the 18 possible states illustrated with black markers in  figure~\ref{Fig:Mtensor_ex}. For each one of these markers, there exists a non-zero two-state tensor $M^{(m,n)}_{ij\; MN}$. Using the same matrix notation as in section~\ref{Sec_ejem}, the two-state tensors involving states with the same degeneracy are:
\begin{eqnarray}
	\mathds{M}^{((n_{1}-4,n_{d}),(n_{1}, n_{d}))} & = & -\frac{\sqrt{n_{1} (n_{1}-1) (n_{1}-2) (n_{1}-3) }}{64 \omega_{1}^{4}}\begin{pmatrix}
		0 & 0 & 0 \\
		0 & 1 & 0 \\
		0 & 0 & 0  
	\end{pmatrix} \Kp \mathds{1}_{d_{n}}, \\
	\mathds{M}^{((n_{1}+4,n_{d}),(n_{1}, n_{d}))} & = & -\frac{\sqrt{(n_{1}+1) (n_{1}+2) (n_{1}+3) (n_{1}+4) }}{64 \omega_{1}^{4}}\begin{pmatrix}
		0 & 0 & 0 \\
		0 & 1 & 0 \\
		0 & 0 & 0  
	\end{pmatrix} \Kp \mathds{1}_{d_{n}},
\end{eqnarray}

\begin{eqnarray}
	\mathds{M}^{((n_{1}-3,n_{d}),(n_{1}, n_{d}))} & = & -\frac{1}{8} \sqrt{\frac{3 n_{1} (n_{1}-1) (n_{1}-2)}{2 \hbar \omega_{1}^{11}}} \begin{pmatrix}
		0 & \omega_{1}^{2} & 0 \\
		\omega_{1}^{2} & -2X & 0 \\
		0 & 0 & 0  
	\end{pmatrix} \Kp \mathds{1}_{d_{n}},\\
	\mathds{M}^{((n_{1}+3,n_{d}),(n_{1}, n_{d}))} & = & -\frac{1}{8} \sqrt{\frac{3 (n_{1}+1) (n_{1}+2) (n_{1}+3)}{2 \hbar \omega_{1}^{11}}} \begin{pmatrix}
		0 & \omega_{1}^{2} & 0 \\
		\omega_{1}^{2} & -2X & 0 \\
		0 & 0 & 0  
	\end{pmatrix} \Kp \mathds{1}_{d_{n}},
\end{eqnarray}

\begin{eqnarray}
	\mathds{M}^{((n_{1}-2,n_{d}),(n_{1}, n_{d}))} & = & - \frac{3 \sqrt{n_{1} (n_{1}-1)}}{ 2 \hbar  \omega_{1}^{7}}\begin{pmatrix}
		\omega_{1}^{4} & - \omega_{1}^{2} X & 0 \\
		- \omega_{1}^{2} X & X^{2} & 0 \\
		0 & 0 & 0  
	\end{pmatrix} \Kp \mathds{1}_{d_{n}}, \\
	\mathds{M}^{((n_{1}+2,n_{d}),(n_{1}, n_{d}))} & = & - \frac{3 \sqrt{( n_{1}+1 ) (n_{1}+2)}}{ 2 \hbar  \omega_{1}^{7}}\begin{pmatrix}
		\omega_{1}^{4} & - \omega_{1}^{2} X & 0 \\
		- \omega_{1}^{2} X & X^{2} & 0 \\
		0 & 0 & 0  
	\end{pmatrix} \Kp \mathds{1}_{d_{n}},
\end{eqnarray}

\begin{eqnarray}
	\mathds{M}^{((n_{1}-1,n_{d}),(n_{1}, n_{d}))} & = & \frac{1}{8} \sqrt{\frac{3 n_{1}}{2 \hbar \omega_{1}^{7}}} \left[ \frac{n_{1}}{\omega_{1}^{2}} \begin{pmatrix}
		0 & \omega_{1}^{2} & 0 \\
		\omega_{1}^{2} & -2X & 0 \\
		0 & 0 & 0  
	\end{pmatrix} + \begin{pmatrix}
		0 & -1 & 0 \\
		1 & 0 & 0 \\
		0 & 0 & 0  
	\end{pmatrix} \right] \Kp \mathds{1}_{d_{n}}, \\
	\mathds{M}^{((n_{1}+1,n_{d}),(n_{1}, n_{d}))} & = & \frac{1}{8} \sqrt{\frac{3 (n_{1}+1)}{2 \hbar \omega_{1}^{7}}} \left[\frac{n_{1}+1}{\omega_{1}^{2}} \begin{pmatrix}
		0 & \omega_{1}^{2} & 0 \\
		\omega_{1}^{2} & -2X & 0 \\
		0 & 0 & 0  
	\end{pmatrix} - \begin{pmatrix}
		0 & -1 & 0 \\
		1 & 0 & 0 \\
		0 & 0 & 0  
	\end{pmatrix} \right] \Kp \mathds{1}_{d_{n}}. \quad
\end{eqnarray}
These tensors are associated with all the black markers along the $n_{1}$ axis in figure~\ref{Fig:Mtensor_ex}. Meanwhile, when the perturbations induce a transition to a state with different degeneracy, we have the case with two-unit jumps, which are on the lines $n_{d}\pm 2$. The two-state tensors for this case are
\begin{eqnarray}
	\mathds{M}^{((n_{1}-2,n_{d}-2),(n_{1}, n_{d}))} & = & -\frac{\sqrt{n_{1} (n_{1}-1)}}{64 \omega_{1}^{2} \omega_{2}^{2}}\begin{pmatrix}
		0 & 0 & 0 \\
		0 & 2 & 3 \\
		0 & 3 & 0  
	\end{pmatrix} \Kp  \mathds{D}(n_{d}) ,\\
	\mathds{M}^{((n_{1}+2,n_{d}+2),(n_{1}, n_{d}))} & = & -\frac{\sqrt{(n_{1} +1)(n_{1}+2)}}{64 \omega_{1}^{2} \omega_{2}^{2}} \begin{pmatrix}
		0 & 0 & 0 \\
		0 & 2 & 3 \\
		0 & 3 & 0  
	\end{pmatrix} \Kp  \mathds{D}^{T}(n_{d}+2) ,
\end{eqnarray}

\begin{eqnarray}
	\mathds{M}^{((n_{1}-2,n_{d}+2),(n_{1}, n_{d}))} & = & \frac{\sqrt{n_{1} (n_{1}-1)}}{64 \omega_{1}^{2} \omega_{2}^{2}}\begin{pmatrix}
		0 & 0 & 0 \\
		0 & 2 & 3 \\
		0 & 3 & 0  
	\end{pmatrix} \Kp  \mathds{D}^{T}(n_{d}+2), \\
	\mathds{M}^{((n_{1}+2,n_{d}-2),(n_{1}, n_{d}))} & = & \frac{\sqrt{(n_{1} +1)(n_{1}+2)}}{64 \omega_{1}^{2} \omega_{2}^{2}} \begin{pmatrix}
		0 & 0 & 0 \\
		0 & 2 & 3 \\
		0 & 3 & 0  
	\end{pmatrix} \Kp  \mathds{D}(n_{d}),
\end{eqnarray}

\begin{eqnarray}
	\mathds{M}^{((n_{1}-1,n_{d}-2),(n_{1}, n_{d}))} & = & -\frac{1}{8 \omega_{2}^{2}} \sqrt{\frac{3 n_{1}}{2 \hbar \omega_{1}^{7}}}\begin{pmatrix}
		0 & \omega_{1}^{2} & 3\omega_{1}^{2} \\
		\omega_{1}^{2} & -2 X & -3 X \\
		3\omega_{1}^{2} & -3 X & 0  
	\end{pmatrix} \Kp  \mathds{D}(n_{d}), \\
	\mathds{M}^{((n_{1}+1,n_{d}+2),(n_{1}, n_{d}))} & = & -\frac{1}{8 \omega_{2}^{2}} \sqrt{\frac{3 (n_{1} + 1)}{2 \hbar \omega_{1}^{7}}}\begin{pmatrix}
		0 & \omega_{1}^{2} & 3\omega_{1}^{2} \\
		\omega_{1}^{2} & -2 X & -3 X \\
		3\omega_{1}^{2} & -3 X & 0  
	\end{pmatrix} \Kp  \mathds{D}^{T}(n_{d}+2),
\end{eqnarray}

\begin{eqnarray}
	\mathds{M}^{((n_{1}-1,n_{d}+2),(n_{1}, n_{d}))} & = & \frac{1}{8 \omega_{2}^{2}} \sqrt{\frac{3 n_{1}}{2 \hbar \omega_{1}^{7}}}\begin{pmatrix}
		0 & \omega_{1}^{2} & 3\omega_{1}^{2} \\
		\omega_{1}^{2} & -2 X & -3 X \\
		3\omega_{1}^{2} & -3 X & 0  
	\end{pmatrix} \Kp  \mathds{D}^{T}(n_{d}+2), \\
	\mathds{M}^{((n_{1}+1,n_{d}-2),(n_{1}, n_{d}))} & = & \frac{1}{8 \omega_{2}^{2}} \sqrt{\frac{3 (n_{1}+1)}{2 \hbar \omega_{1}^{7}}}\begin{pmatrix}
		0 & \omega_{1}^{2} & 3\omega_{1}^{2} \\
		\omega_{1}^{2} & -2 X & -3 X \\
		3\omega_{1}^{2} & -3 X & 0  
	\end{pmatrix} \Kp  \mathds{D}(n_{d}).
\end{eqnarray}
Finally, the last tensors arise when we change to a state with degeneracy of $n_{d}\pm 4$ with respect to the initial state. The two-state tensors in this situation are 
\begin{eqnarray}
	\mathds{M}^{((n_{1},n_{d}-4),(n_{1}, n_{d}))} & = & -\frac{1}{64 \omega_{2}^{4}}\begin{pmatrix}
		0 & 0 & 0 \\
		0 & 1 & 3 \\
		0 & 3 & 9  
	\end{pmatrix} \Kp \left[ \mathds{D}(n_{d}-2) \cdot \mathds{D}(n_{d})  \right], \\
	\mathds{M}^{((n_{1},n_{d}+4),(n_{1}, n_{d}))} & = & -\frac{1}{64 \omega_{2}^{4}}\begin{pmatrix}
		0 & 0 & 0 \\
		0 & 1 & 3 \\
		0 & 3 & 9  
	\end{pmatrix} \Kp  \left[ \mathds{D}^{T}(n_{d}+4) \cdot \mathds{D}^{T}(n_{d}+2)  \right].
\end{eqnarray}

Of all the 18 tensors, only $\mathds{M}^{((n_{1}\pm 1, n_{d}),(n_{1}, n_{d}))}$ has an anti-symmetric part. Therefore, in all the others $\mathds{M}^{((m_{1}, m_{d}),(n_{1}, n_{d}))}=\mathds{G}^{((m_{1}, m_{d}),(n_{1}, n_{d}))}$, and the order of the perturbations does not matter. In contrast, for $(m_{1}, m_{d})=(n_{1}\pm 1, n_{d})$ the symmetric part of the two-state tensors are
\begin{eqnarray}
	\mathds{G}^{((n_{1}-1,n_{d}),(n_{1}, n_{d}))} & = & \frac{n_{1}}{8} \sqrt{\frac{3 n_{1}}{2 \hbar \omega_{1}^{11}}}   \begin{pmatrix}
		0 & \omega_{1}^{2} & 0 \\
		\omega_{1}^{2} & -2X & 0 \\
		0 & 0 & 0  
	\end{pmatrix}  \Kp \mathds{1}_{d_{n}}, \\
	\mathds{G}^{((n_{1}+1,n_{d}),(n_{1}, n_{d}))} & = & \frac{n_{1}+1}{8} \sqrt{\frac{3 (n_{1}+1)}{2 \hbar \omega_{1}^{11}}}  \begin{pmatrix}
		0 & \omega_{1}^{2} & 0 \\
		\omega_{1}^{2} & -2X & 0 \\
		0 & 0 & 0  
	\end{pmatrix} \Kp \mathds{1}_{d_{n}}. 
\end{eqnarray}
Meanwhile, the torsion, proportional to the anti-symmetric part of $\mathds{M}^{((n_{1}\pm 1, n_{d}),(n_{1}, n_{d}))}$, is
\begin{eqnarray}
	\mathds{T}^{((n_{1}-1,n_{d}),(n_{1}, n_{d}))} & = & \frac{\mathrm{i}}{4} \sqrt{\frac{3 n_{1}}{2 \hbar \omega_{1}^{7}}}  \begin{pmatrix}
		0 & -1 & 0 \\
		1 & 0 & 0 \\
		0 & 0 & 0  
	\end{pmatrix} \Kp \mathds{1}_{d_{n}}, \\
	\mathds{T}^{((n_{1}+1,n_{d}),(n_{1}, n_{d}))} & = & - \frac{\mathrm{i}}{4} \sqrt{\frac{3 (n_{1}+1)}{2 \hbar \omega_{1}^{7}}}  \begin{pmatrix}
		0 & -1 & 0 \\
		1 & 0 & 0 \\
		0 & 0 & 0  
	\end{pmatrix}  \Kp \mathds{1}_{d_{n}}. 
\end{eqnarray}
All the two-state tensors with jumps that preserve the degeneracy are multiplied (via the Kronecker product) by a $d_{n}\times d_{n}$ identity matrix. In these cases, the direction within the new subspace $\mathcal{H}_{m}$ remains unchanged ($M=N$). Conversely, when there is a change in the degeneracy, the $N$ direction may also change. In this situation, the tensors are multiplied by a non-diagonal matrix. If the matrix is $\mathds{D}(n_{d})$, the system either remains in the same direction ($M=N$) or shifts two units down ($M=N-2$). Meanwhile, if the matrix is $\mathds{D}^{T}(n_{d}+2)$, one may again stay in the same direction ($M=N$) or move two units up ($M=N+2$). Finally, the tensor $\mathds{M}^{((n_{1},n_{d}-4),(n_{1}, n_{d}))}$ is multiplied by $\mathds{D}(n_{d}-2) \cdot \mathds{D}(n_{d})$, while $\mathds{M}^{((n_{1},n_{d} + 4),(n_{1}, n_{d}))}$ by $\mathds{D}^{T}(n_{d}+4) \cdot \mathds{D}^{T}(n_{d}+2)$. The first product yields a $(d_{n}-4)\times d_{n}$ matrix that allows transitions to $M=N,N-2,N-4$. The second gives a $(d_{n}+4)\times d_{n}$ matrix that permits transitions to $M=N,N+2,N+4$. Therefore, as we move further to states with more (or less) degeneration, we enable more transitions in the directions within the new subspace $\mathcal{H}_{m}$.

\subsection{Invariants}

Let us end this section with a discussion on the invariants. The summands of the QGT $q^{(n,m)}_{ij}$ are the first example of two-state invariants. They represent a close path due to two parameter variations. The path starts and ends at the state $n$ after it passed through a different state $m$. Although we did not show them for the example in section~\ref{Sec_ejem}, they are:
\begin{subequations}
	\begin{eqnarray}
		\mfs{q^{((n_{1}, n_{d}),(n_{1}-1,n_{d}))}_{ij}  } & = & \frac{3n_{1}(n_{d}+1)}{2 \hbar \omega_{1}^{7}}\begin{pmatrix}
			\omega_{1}^{4} & - \omega_{1}^{2} X & 0 \\
			- \omega_{1}^{2} X & X^{2} & 0 \\
			0 & 0 & 0  
		\end{pmatrix}, \\
		\mfs{q^{((n_{1}, n_{d}),(n_{1}+1,n_{d}))}_{ij}  } & = & \frac{3(n_{1}+1)(n_{d}+1)}{2 \hbar \omega_{1}^{7}}\begin{pmatrix}
			\omega_{1}^{4} & - \omega_{1}^{2} X & 0 \\
			- \omega_{1}^{2} X & X^{2} & 0 \\
			0 & 0 & 0  
		\end{pmatrix}, \\
		\mfs{q^{((n_{1}, n_{d}),(n_{1}-2,n_{d}))}_{ij}  } & = & \frac{n_{1} ( n_{1} - 1)(n_{d} + 1)}{32 \omega_{1}^{4}}\begin{pmatrix}
			0 & 0 & 0 \\
			0 & 1 & 0 \\
			0 & 0 & 0  
		\end{pmatrix}, \\
		\mfs{q^{((n_{1}, n_{d}),(n_{1}+2,n_{d}))}_{ij}  } & = & \frac{(n_{1} + 1 ) (n_{1} + 2)(n_{d} + 1)}{32 \omega_{1}^{4}}\begin{pmatrix}
			0 & 0 & 0 \\
			0 & 1 & 0 \\
			0 & 0 & 0  
		\end{pmatrix}, \\
		\mfs{q^{((n_{1}, n_{d}),(n_{1},n_{d}-2))}_{ij}  } & = & \frac{n_{d}(n_{d}^{2}-1)}{96 \omega_{2}^{4}}\begin{pmatrix}
			0 & 0 & 0 \\
			0 & 1 & 3 \\
			0 & 3 & 9  
		\end{pmatrix}, \\
		\mfs{q^{((n_{1}, n_{d}),(n_{1},n_{d}+2))}_{ij}  } & = & \frac{(n_{d}+1)(n_{d}+2)(n_{d}+3)}{96 \omega_{2}^{4}}\begin{pmatrix}
			0 & 0 & 0 \\
			0 & 1 & 3 \\
			0 & 3 & 9  
		\end{pmatrix}.
	\end{eqnarray}
\end{subequations}
All of them are real and therefore symmetric. Thus, they qualify as observables. Moreover, the sum over all intermediate states $m=(m_{1}, m_{d})$ yields the trace of QGT
\begin{eqnarray}
	Q^{(n_{1},n_{d})}_{ij} = \sum_{N=1}^{d_{n}} Q^{(n_{1},n_{d})}_{ij\; NN} &=& q^{((n_{1}, n_{d}),(n_{1}-1,n_{d}))}_{ij} + q^{((n_{1}, n_{d}),(n_{1}+1,n_{d}))}_{ij} + q^{((n_{1}, n_{d}),(n_{1}-2,n_{d}))}_{ij} \nonumber \\
	&& + q^{((n_{1}, n_{d}),(n_{1}+2,n_{d}))}_{ij} + q^{((n_{1}, n_{d}),(n_{1},n_{d}-2))}_{ij} + q^{((n_{1}, n_{d}),(n_{1},n_{d}+2))}_{ij}. \quad
\end{eqnarray}
The observables $q^{(n,m)}_{ij}$ depend on the initial energy level $n=(n_{1}, n_{d})$ and on the parameters $\lambda=\{X,k,k^{\prime}\}$. Also, the matrix structure encodes the path required to return to the initial state. For example, to reach the state $n=(n_{1}, n_{d})$ after passing through $m=(n_{1}-1, n_{d})$, we either vary the $X$ or $k$ parameter twice, or we apply a combination of them ($X$ and then $k$ or vice versa). The matrix symmetry ensures that the order in which we perform the variations is irrelevant. Furthermore, no variation on $k^{\prime}$ allows us to pass through $m=(n_{1}-1, n_{d})$. In contrast, to pass through the state $m=(n_{1}, n_{d}-2)$, no variation on the field strength $X$ is required. 

Considering the scalar invariants of Figs.~\ref{InvN_1} and \ref{InvN_2}, the only state $m$ for which the three tensors do not vanish are $m=(n_{1}+1, n_{d})$. In this case, the invariants are:
\begin{subequations}
	\begin{eqnarray}
		\mathcal{N}_{M}^{((n_{1}+1,n_{d}),(n_{1},n_{d}))}&=&\frac{2(n_{1}+1)^{2}(2n_{1}^{4}+8n_{1}^{3}+17n_{1}^{2}+18n_{1}+6)}{(2n_{1}+1)(2n_{1}+3)(n_{1}^{2}+n_{1}+1)(n_{1}^{2}+3n_{1}+3)(n_{d}+1)}, \\
		\mathcal{N}_{\mathcal{G}}^{((n_{1}+1,n_{d}),(n_{1},n_{d}))}&=&\frac{2(n_{1}+1)^{4}(2n_{1}^{2}+4n_{1}+3)}{(2n_{1}+1)(2n_{1}+3)(n_{1}^{2}+n_{1}+1)(n_{1}^{2}+3n_{1}+3)(n_{d}+1)},\\
		\mathcal{N}_{T}^{((n_{1}+1,n_{d}),(n_{1},n_{d}))} & = &- \frac{8(n_{1}+1)^{2}(2n_{1}^{2}+4n_{1}+3)}{(2n_{1}+1)(2n_{1}+3)(n_{1}^{2}+n_{1}+1)(n_{1}^{2}+3n_{1}+3)(n_{d}+1)}. 
	\end{eqnarray}
\end{subequations}
Together with 
\begin{equation}
	\mathcal{N}_{\mathcal{G}T}^{((n_{1}+1,n_{d}),(n_{1},n_{d}))} = - \frac{4 \mathrm{i} n_{1} (n_{1}+1)^{2} (n_{1}+2)}{(2n_{1}+1)(2n_{1}+3)(n_{1}^{2}+n_{1}+1)(n_{1}^{2}+3n_{1}+3)(n_{d}+1)},
\end{equation}
they satisfy the identity
\begin{equation}
	\mathcal{N}_{M}^{((n_{1}+1,n_{d}),(n_{1},n_{d}))} =  \mathcal{N}_{\mathcal{G}}^{((n_{1}+1,n_{d}),(n_{1},n_{d}))} - \frac{1}{4} \mathcal{N}_{T}^{((n_{1}+1,n_{d}),(n_{1},n_{d}))} + \mathrm{Im} \left\lbrace \mathcal{N}_{\mathcal{G}T}^{((n_{1}+1,n_{d}),(n_{1},n_{d}))} \right\rbrace.
\end{equation}
For the other states, we do not have torsion. Hence, $\mathcal{N}^{(n,m)}_{M}=\mathcal{N}^{(n,m)}_{\mathcal{G}}$ and $\mathcal{N}^{(n,m)}_{\mathcal{G}T}=0$. Meanwhile, completing the scheme for the invariants whose degeneracy does not change, we have:	
\begin{subequations}
	\begin{eqnarray}
		\mathcal{N}_{M}^{((n_{1}+2,n_{d}),(n_{1},n_{d}))}&=&\frac{2(n_{1}+1)(n_{1}+2)}{(2n_{1}+1)(2n_{1}+5)(n_{d}+1)},\\
		\mathcal{N}_{M}^{((n_{1}+3,n_{d}),(n_{1},n_{d}))}&=&\frac{2(n_{1}+1)(n_{1}+2)^{2}(n_{1}+3)(2n_{1}^{2}+8n_{1}+5)}{(2n_{1}+1)(2n_{1}+7)(n_{1}^{2}+n_{1}+1)(n_{1}^{2}+7n_{1}+13)(n_{d}+1)},\\
		\mathcal{N}_{M}^{((n_{1}+4,n_{d}),(n_{1},n_{d}))}&=&\frac{(n_{1}+1)(n_{1}+2)(n_{1}+3)(n_{1}+4)}{2(n_{1}^{2}+n_{1}+1)[(n_{1}+4)^{2}+n_{1}+5](n_{d}+1)}.
	\end{eqnarray}
\end{subequations}
All of this cases depend on the degeneracy number $n_{d}$ as $(n_{d}+1)^{-1}$. Therefore, the structure of these invariants scales down as $n_{d}$ grows. Furthermore, for a given degeneracy, when $n_{1}$ tends to infinity, the invariants converge to three possible scenarios. Either they depend on the degeneracy:
\begin{equation}
	\lim_{n_{1}\to \infty} \mathcal{N}_{M}^{((n_{1} + 1,n_{d}),(n_{1},n_{d}))}=\lim_{n_{1}\to \infty} \mathcal{N}_{\mathcal{G}}^{((n_{1} + 1,n_{d}),(n_{1},n_{d}))}=\lim_{n_{1}\to \infty} \mathcal{N}_{M}^{((n_{1}+ 3,n_{d}),(n_{1},n_{d}))}=\frac{1}{1+n_{d}}
\end{equation}
and
\begin{equation}
	\lim_{n_{1}\to \infty} \mathcal{N}_{M}^{((n_{1} + 2,n_{d}),(n_{1},n_{d}))}=\lim_{n_{1}\to \infty} \mathcal{N}_{M}^{((n_{1} + 4,n_{d}),(n_{1},n_{d}))}=\frac{1}{2}\left(\frac{1}{1+n_{d}}\right),
\end{equation}
or they approach to zero
\begin{equation}
	\lim_{n_{1}\to \infty} \mathcal{N}_{T}^{((n_{1} + 1,n_{d}),(n_{1},n_{d}))}=0.
\end{equation}
These invariants that do not change the degeneracy are equivalent to those in~\cite{Romero2409}, with the key difference being the different convention they used for the scalar invariant, see discussion in appendix~\ref{Ap_Sec_inv}. In particular, $\mathcal{N}_{T}^{((n_{1} + 1,n_{d}),(n_{1},n_{d}))}$ is positive in their convention, so the invariant $\mathcal{N}_{M}^{((n_{1} + 1,n_{d}),(n_{1},n_{d}))}$ is also different. Nonetheless, both results are compatible in the sense that they share the same functional form.

The next invariants are those that relate states with different values of $n_{1}$ and degeneracy $n_{d}$. In this case, we have the invariants that change by one unit in $n_{1}$
\begin{subequations}
	\begin{eqnarray}
		\mathcal{N}_{M}^{((n_{1}-1,n_{d}+2),(n_{1},n_{d}))}&=&\frac{4n_{1}(n_{d}+2)[n_{1}(n_{d}^{2}+4n_{d}+7)+n_{d}+2]}{(2n_{1}+1)(2n_{1}-1)(n_{d}^{2}+2n_{d}+3)(n_{d}^{2}+6n_{d}+11)},\\
		\mathcal{N}_{M}^{((n_{1}+1,n_{d}+2),(n_{1},n_{d}))}&=&\frac{4(n_{1}+1)(n_{d}+2)[(n_{1}+1)(n_{d}^{2}+4n_{d}+7)-(n_{d}+2)]}{(2n_{1}+1)(2n_{1}+3)(n_{d}^{2}+2n_{d}+3)(n_{d}^{2}+6n_{d}+11)},  
	\end{eqnarray}
\end{subequations}       
and those that change by two units in $n_{1}$
\begin{subequations}
	\begin{eqnarray}
		\mathcal{N}_{M}^{((n_{1}-2,n_{d}+2),(n_{1},n_{d}))}&=&\frac{n_{1}(n_{1}-1)(n_{d}+2)[(n_{1}^{2}-n_{1}+2)(n_{d}^{2}+4n_{d}+7)+2(2n_{1}-1)(n_{d}+2)]}{(n_{1}^{2}-3n_{1}+3)(n_{1}^{2}+n_{1}+1)(n_{d}^{2}+2n_{d}+3)(n_{d}^{2}+6n_{d}+11)},\nonumber \\ \\
		\mathcal{N}_{M}^{((n_{1}+2,n_{d}+2),(n_{1},n_{d}))}&=&\frac{(n_{1}+1)(n_{1}+2)(n_{d}+2)}{(n_{1}^{2}+n_{1}+1)(n_{1}^{2}+5n_{1}+7)(n_{d}^{2}+2n_{d}+3)(n_{d}^{2}+6n_{d}+11)}\\
		& &\times\left[(n_{1}^{2}+3n_{1}+4)(n_{d}^{2}+4n_{d}+7) \right.\nonumber
		\left. -2(2n_{1}+3)(n_{d}+2)\right].
	\end{eqnarray}
\end{subequations}
In both cases, the dependency on $n_{d}$ is more complicated. Nonetheless, as $n_{1}$ increases its value, all of them converge to the same function:
\begin{equation}
	\label{n1_limit2}
	\lim_{n_{1}\to \infty} \mathcal{N}_{M}^{((n_{1}\pm1,n_{d}+2),(n_{1},n_{d}))}=\lim_{n_{1}\to \infty} \mathcal{N}_{M}^{((n_{1}\pm2,n_{d}+2),(n_{1},n_{d}))}=\frac{n_{d}^{3}+6n_{d}^{2}+15n_{d}+14}{n_{d}^{4}+8n_{d}^{3}+26n_{d}^{2}+40n_{d}+33}.
\end{equation}
Therefore, for $n_{1}$ big enough, the jumps of one or two units in $n_{1}$ are not relevant, and all the invariants of this type behave the same.  Finally, we have a two-state geometric tensor that relates states only with a difference in $n_{d}$. Here, $m=(n_{1}, n_{d}+4)$, and its scalar invariant only depends on $n_{d}$. Thus, it appears as a constant line on figure~\ref{InvN_2}. It is the only one of this type; explicitly, it is
\begin{eqnarray}
	\label{NMndp4}
	\mathcal{N}_{M}^{((n_{1},n_{d}+4),(n_{1},n_{d}))}&=&\frac{3(n_{d}^{3}+9n_{d}^{2}+26n_{d}+24)}{5(n_{d}^{2}+2n_{d}+3)(n_{d}^{2}+10n_{d}+27)}.
\end{eqnarray}

\begin{figure}[h]%
	\centering
	{\includegraphics[scale=0.9]{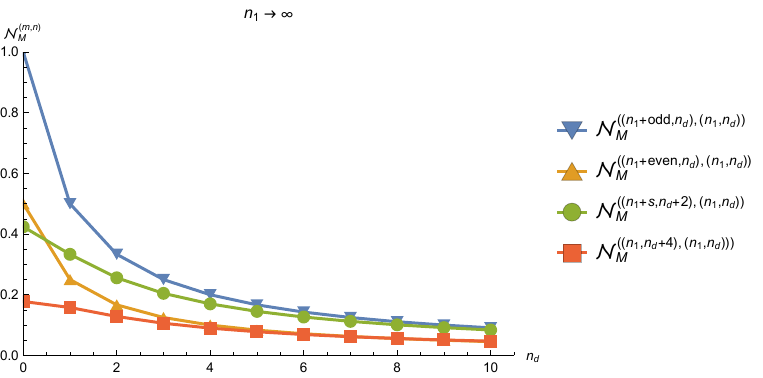}} 
	\caption{Plot showing the four different values of the invariants when $n_{1}\rightarrow \infty$. Here, $s=\pm 1,\, \pm 2$.}
	\label{InvN_lim}
\end{figure}

Analyzing the invariants for sufficiently large $n_{1}$ reveals four distinct scenarios (excluding the torsion invariant, which vanishes for all $n_{d}$). The first two correspond to transitions that preserve the degeneracy, with odd or even increments in $n_{1}$,  $\mathcal{N}_{M}^{((n_{1} + \text{odd},n_{d}),(n_{1},n_{d}))}$ and $\mathcal{N}_{M}^{((n_{1} + \text{even}, n_{d}),(n_{1},n_{d}))}$. The third scenario involves a two-unit change in the degeneracy accompanied by a shift of one or two units in $n_{1}$, $\mathcal{N}_{M}^{((n_{1} + s, n_{d}+2),(n_{1},n_{d}))}$, with $s=\pm 1,\, \pm 2$. Finally, the fourth case corresponds to a transition towards a state four units more degenerate, which is always independent of $n_{1}$.  As shown in figure~\ref{InvN_lim}, these cases yield distinctive invariant values near $n_{d}=0$. However, as $n_{d}$ increases, all of them converge to zero. Combining these results with those at $n_{d}=0$ in Figs.~\ref{InvN_1} and \ref{InvN_2}, the invariants show more sensitivity in the states close to $n_{1}=0$ or $n_{d}=0$. This sensitivity suggests that measurements of these quantities may be most feasible near the ground state of the quantum system.

The trace of the QGT, its summands, and the vector invariant \eqref{inv_vec_ex} depend on the parameters that characterize the quantum system. Consequently, they are susceptible to the geometry of the parameter space. In contrast, the scalar invariants shown in Figs.~\ref{InvN_1} and \ref{InvN_2} are independent of the parameters and rely solely on the energy levels. Thus, the nature of these invariants is determined exclusively by the topology of subspaces $\mathcal{H}_{n}$ and $\mathcal{H}_{m}$.

\end{document}